\documentclass{ws-mpla}
\usepackage[super]{cite}
\usepackage{graphicx}
\usepackage{xcolor}
\usepackage{amsmath}
\usepackage{tikz}
\usepackage{tikz-feynman}
\usepackage{hyperref}
\usepackage{enumitem}

\def\ubar{\overline{u}}

\def\lbar{\overline{\ell}}

\def\Lbar{\overline{L}}

\def\invfb   {\ensuremath{\mbox{\,fb}^{-1}}\xspace}

\def\invab   {\ensuremath{\mbox{\,ab}^{-1}}\xspace}

\newcommand{\tev}{{\ensuremath{\mathrm{\,Te\kern -0.1em V}}}\xspace}
\newcommand{\gev}{\ensuremath{\mathrm{\,Ge\kern -0.1em V}}\xspace}
\newcommand{\mev}{\ensuremath{\mathrm{\,Me\kern -0.1em V}}\xspace}
\newcommand{\kev}{\ensuremath{\mathrm{\,ke\kern -0.1em V}}\xspace}
\newcommand{\ev}{\ensuremath{\mathrm{\,e\kern -0.1em V}}\xspace}
\newcommand{\gevc}{\ensuremath{{\mathrm{\,Ge\kern -0.1em V\!/}c}}\xspace}
\newcommand{\mevc}{\ensuremath{{\mathrm{\,Me\kern -0.1em V\!/}c}}\xspace}
\newcommand{\gevcc}{\ensuremath{{\mathrm{\,Ge\kern -0.1em V\!/}c^2}}\xspace}
\newcommand{\gevgevcccc}{\ensuremath{{\mathrm{\,Ge\kern -0.1em V^2\!/}c^4}}\xspace}
\newcommand{\mevcc}{\ensuremath{{\mathrm{\,Me\kern -0.1em V\!/}c^2}}\xspace}

\usepackage{xspace} 
\usepackage{upgreek}
\def\lhcb {\mbox{LHCb}\xspace}

 \def\Peta        {\ensuremath{\upeta}\xspace}

 \def\Pmu         {\ensuremath{\upmu}\xspace}

 \def\Ppi         {\ensuremath{\uppi}\xspace}                 
                  
 \def\Prho        {\ensuremath{\uprho}\xspace}

 \def\Pomega      {\ensuremath{\upomega}\xspace}                 

 \def\PDelta      {\ensuremath{\Delta}\xspace}                 
 \def\PXi      {\ensuremath{\Xi}\xspace}                 
 \def\PLambda      {\ensuremath{\Lambda}\xspace}                 
 \def\PSigma      {\ensuremath{\Sigma}\xspace}                 
 \def\POmega      {\ensuremath{\Omega}\xspace}                 
 \def\PUpsilon      {\ensuremath{\Upsilon}\xspace}

\def\lhcb {\mbox{LHCb}\xspace}

\def\babar  {\mbox{BaBar}\xspace}
\def\belle  {\mbox{Belle}\xspace}
\def\cleo2  {\mbox{CLEO II}\xspace}
\def\bes3  {\mbox{BES III}\xspace}
\def\E791  {\mbox{E791}\xspace}
\def\E653  {\mbox{E653}\xspace}
\def\cleo   {\mbox{CLEO}\xspace}

\def\argus  {\mbox{ARGUS}\xspace}

\def\CP                {{\ensuremath{C\!P}}\xspace}

\def\BF         {{\ensuremath{\mathcal{B}}}\xspace}

 \def\PB      {\ensuremath{\mathrm{B}}\xspace}                 
                  
 \def\PD      {\ensuremath{\mathrm{D}}\xspace}

 \def\PK      {\ensuremath{\mathrm{K}}\xspace}

 \def\Pb      {\ensuremath{\mathrm{b}}\xspace}                 
 \def\Pc      {\ensuremath{\mathrm{c}}\xspace}                 
 \def\Pd      {\ensuremath{\mathrm{d}}\xspace}                 
 \def\Pe      {\ensuremath{\mathrm{e}}\xspace}

 \def\Pi      {\ensuremath{\mathrm{i}}\xspace}

 \def\Pp      {\ensuremath{\mathrm{p}}\xspace}

 \def\Ps      {\ensuremath{\mathrm{s}}\xspace}

 \def\Peta        {\ensuremath{\eta}\xspace}

 \def\Pmu         {\ensuremath{\mu}\xspace}

 \def\Ppi         {\ensuremath{\pi}\xspace}                 
                  
 \def\Prho        {\ensuremath{\rho}\xspace}

 \def\Pomega      {\ensuremath{\omega}\xspace}                 
 \mathchardef\PDelta="7101
 \mathchardef\PXi="7104
 \mathchardef\PLambda="7103
 \mathchardef\PSigma="7106
 \mathchardef\POmega="710A
 \mathchardef\PUpsilon="7107
                  
 \def\PB      {\ensuremath{B}\xspace}                 
                  
 \def\PD      {\ensuremath{D}\xspace}

 \def\PK      {\ensuremath{K}\xspace}

 \def\Pb      {\ensuremath{b}\xspace}                 
 \def\Pc      {\ensuremath{c}\xspace}                 
 \def\Pd      {\ensuremath{d}\xspace}                 
 \def\Pe      {\ensuremath{e}\xspace}

 \def\Pi      {\ensuremath{i}\xspace}

 \def\Pp      {\ensuremath{p}\xspace}

 \def\Ps      {\ensuremath{s}\xspace}

\def\en         {{\ensuremath{\Pe^-}}\xspace}   
\def\ep         {{\ensuremath{\Pe^+}}\xspace}
\def\epm        {{\ensuremath{\Pe^\pm}}\xspace} 
\def\emp        {{\ensuremath{\Pe^\mp}}\xspace} 
\def\epem       {{\ensuremath{\Pe^+\Pe^-}}\xspace}

\def\mupm        {{\ensuremath{\Pmu^\pm}}\xspace} 
\def\mump        {{\ensuremath{\Pmu^\mp}}\xspace} 

\def\mup        {{\ensuremath{\Pmu^+}}\xspace}
\def\mun        {{\ensuremath{\Pmu^-}}\xspace} 
\def\mumu       {{\ensuremath{\Pmu^+\Pmu^-}}\xspace}

\def\dquark    {{\ensuremath{\Pd}}\xspace}

\def\squark    {{\ensuremath{\Ps}}\xspace}

\def\cquark    {{\ensuremath{\Pc}}\xspace}

\def\bquark    {{\ensuremath{\Pb}}\xspace}

\def\pion   {{\ensuremath{\Ppi}}\xspace}
\def\piz    {{\ensuremath{\pion^0}}\xspace}

\def\pip    {{\ensuremath{\pion^+}}\xspace}
\def\pim    {{\ensuremath{\pion^-}}\xspace}

\def\rhomeson {{\ensuremath{\Prho}}\xspace}
\def\rhoz     {{\ensuremath{\rhomeson^0}}\xspace}
\def\rhop     {{\ensuremath{\rhomeson^+}}\xspace}
\def\rhom     {{\ensuremath{\rhomeson^-}}\xspace}

\def\kaon    {{\ensuremath{\PK}}\xspace}
  \def\Kbar    {{\kern 0.2em\overline{\kern -0.2em \PK}{}}\xspace}

\def\KorKbar    {\kern 0.18em\optbar{\kern -0.18em K}{}\xspace}

\def\Kzb     {{\ensuremath{\Kbar{}^0}}\xspace}
\def\Kp      {{\ensuremath{\kaon^+}}\xspace}
\def\Km      {{\ensuremath{\kaon^-}}\xspace}

\def\KS      {{\ensuremath{\kaon^0_{\mathrm{ \scriptscriptstyle S}}}}\xspace}

\def\Kstarz  {{\ensuremath{\kaon^{*0}}}\xspace}
\def\Kstarzb {{\ensuremath{\Kbar{}^{*0}}}\xspace}

\def\Kstarp  {{\ensuremath{\kaon^{*+}}}\xspace}
\def\Kstarm  {{\ensuremath{\kaon^{*-}}}\xspace}

\newcommand{\etapr}{\ensuremath{\Peta^{\prime}}\xspace}

\newcommand{\omegaz}{\ensuremath{\Pomega}\xspace}

  \def\Dbar    {{\kern 0.2em\overline{\kern -0.2em \PD}{}}\xspace}
\def\D       {{\ensuremath{\PD}}\xspace}

\def\DorDbar    {\kern 0.18em\optbar{\kern -0.18em D}{}\xspace}
\def\Dz      {{\ensuremath{\D^0}}\xspace}
\def\Dzb     {{\ensuremath{\Dbar{}^0}}\xspace}
\def\Dp      {{\ensuremath{\D^+}}\xspace}

\def\Dm      {{\ensuremath{\D^-}}\xspace}

\def\Dstarp  {{\ensuremath{\D^{*+}}}\xspace}
\def\Dstarm  {{\ensuremath{\D^{*-}}}\xspace}

\def\Ds      {{\ensuremath{\D^+_\squark}}\xspace}
\def\Dsp     {{\ensuremath{\D^+_\squark}}\xspace}

\def\B       {{\ensuremath{\PB}}\xspace}
\def\Bbar    {{\ensuremath{\kern 0.18em\overline{\kern -0.18em \PB}{}}}\xspace}

\def\BorBbar    {\kern 0.18em\optbar{\kern -0.18em B}{}\xspace}

\def\Bd      {{\ensuremath{\B^0}}\xspace}

  \def\Y#1S{\ensuremath{\PUpsilon{(#1S)}}\xspace}

\def\proton      {{\ensuremath{\Pp}}\xspace}

\def\Lbar        {{\ensuremath{\kern 0.1em\overline{\kern -0.1em\PLambda}}}\xspace}
\def\LorLbar    {\kern 0.18em\optbar{\kern -0.18em \PLambda}{}\xspace}

\newcommand{\Dkkmm}{\mbox{\ensuremath{\Dz\to\Kp\Km\mu^+\mu^-}}\xspace}
\newcommand{\Dppmm}{\mbox{\ensuremath{\Dz\to\pip\pim\mu^+\mu^-}}\xspace}
\newcommand{\Dkpmm}{\ensuremath{\Dz\to\Km\pip[\mu^+\mu^-]_{\rhoz/\omegaz}}\xspace}
\newcommand{\Dkpee}{\ensuremath{\Dz\to\Km\pip[e^+e^-]_{\rhoz/\omegaz}}\xspace}

\newcommand{\Dmm}{\ensuremath{\Dz \to \mup \mun}\xspace}
\newcommand{\Dee}{\ensuremath{\Dz\to \ep \en}\xspace}
\newcommand{\Demu}{\ensuremath{\Dz\to \mupm \emp}\xspace}

\newcommand{\mmumu}{\ensuremath{m(\mup\mun)}\xspace}

\newcommand{\Acp}{\ensuremath{A_{\CP}}\xspace}

\newcommand{\Afb}{\ensuremath{A_{\mathrm{FB}}}\xspace}
\newcommand{\Aphi}{\ensuremath{A_{2\phi}}\xspace}

\newcommand{\Acpppmm}{\ensuremath{4.9}\xspace}
\newcommand{\AcpppmmStat}{\ensuremath{3.8}\xspace}
\newcommand{\AcpppmmSyst}{\ensuremath{0.7}\xspace}
\newcommand{\Afbppmm}{\ensuremath{3.3}\xspace}
\newcommand{\AfbppmmStat}{\ensuremath{3.7}\xspace}
\newcommand{\AfbppmmSyst}{\ensuremath{0.6}\xspace}
\newcommand{\Aphippmm}{\ensuremath{-0.6}\xspace}
\newcommand{\AphippmmStat}{\ensuremath{3.7}\xspace}
\newcommand{\AphippmmSyst}{\ensuremath{0.6}\xspace}
\newcommand{\Acpkkmm}{\ensuremath{0}\xspace}
\newcommand{\AcpkkmmStat}{\ensuremath{11}\xspace}
\newcommand{\AcpkkmmSyst}{\ensuremath{2}\xspace}
\newcommand{\Afbkkmm}{\ensuremath{0}\xspace}
\newcommand{\AfbkkmmStat}{\ensuremath{11}\xspace}
\newcommand{\AfbkkmmSyst}{\ensuremath{2}\xspace}
\newcommand{\Aphikkmm}{\ensuremath{9}\xspace}
\newcommand{\AphikkmmStat}{\ensuremath{11}\xspace}
\newcommand{\AphikkmmSyst}{\ensuremath{1}\xspace}

\newcommand{\BFppmm}{\ensuremath{9.64}\xspace}
\newcommand{\BFppmmStat}{\ensuremath{0.48}\xspace}
\newcommand{\BFppmmSyst}{\ensuremath{0.51}\xspace}
\newcommand{\BFppmmNorm}{\ensuremath{0.97}\xspace}
\newcommand{\BFppmmUnit}{\ensuremath{\times10^{-7}}\xspace}
\newcommand{\BFkkmm}{\ensuremath{1.54}\xspace}
\newcommand{\BFkkmmStat}{\ensuremath{0.27}\xspace}
\newcommand{\BFkkmmSyst}{\ensuremath{0.09}\xspace}
\newcommand{\BFkkmmNorm}{\ensuremath{0.16}\xspace}
\newcommand{\BFkkmmUnit}{\ensuremath{\times10^{-7}}\xspace}
\newcommand{\BFkpmm}{\ensuremath{4.17}\xspace}
\newcommand{\BFkpmmStat}{\ensuremath{0.12}\xspace}
\newcommand{\BFkpmmSyst}{\ensuremath{0.40}\xspace}

\newcommand{\BFkpmmUnit}{\ensuremath{\times10^{-6}}\xspace}

\newcommand{\BFkpee}{\ensuremath{4.0}\xspace}
\newcommand{\BFkpeeStat}{\ensuremath{0.5}\xspace}
\newcommand{\BFkpeeSyst}{\ensuremath{0.2}\xspace}
\newcommand{\BFkpeeNorm}{\ensuremath{0.1}\xspace}

\newcommand{\BFkpeeUnit}{\ensuremath{\times10^{-6}}\xspace}

\newcommand{\BFrg}{\ensuremath{1.77}\xspace}
\newcommand{\BFrgStat}{\ensuremath{0.30}\xspace}
\newcommand{\BFrgSyst}{\ensuremath{0.07}\xspace}
\newcommand{\BFrgUnit}{\ensuremath{\times10^{-5}}\xspace}

\newcommand{\BFkg}{\ensuremath{4.66}\xspace}
\newcommand{\BFkgStat}{\ensuremath{0.21}\xspace}
\newcommand{\BFkgSyst}{\ensuremath{0.21}\xspace}
\newcommand{\BFkgUnit}{\ensuremath{\times10^{-4}}\xspace}

\newcommand{\BFpg}{\ensuremath{2.76}\xspace}
\newcommand{\BFpgStat}{\ensuremath{0.19}\xspace}
\newcommand{\BFpgSyst}{\ensuremath{0.10}\xspace}
\newcommand{\BFpgUnit}{\ensuremath{\times10^{-6}}\xspace}

\newcommand{\Dzrg}{\ensuremath{\Dz \to \rhoz \gamma}\xspace}
\newcommand{\Dzkg}{\ensuremath{\Dz \to \Kstarzb \gamma}\xspace}
\newcommand{\Dzpg}{\ensuremath{\Dz \to \phi \gamma}\xspace}

\newcommand{\Acprg}{\ensuremath{5.6}\xspace}
\newcommand{\AcprgStat}{\ensuremath{15.2}\xspace}
\newcommand{\AcprgSyst}{\ensuremath{0.6}\xspace}

\newcommand{\Acpkg}{\ensuremath{-0.3}\xspace}
\newcommand{\AcpkgStat}{\ensuremath{2.0}\xspace}
\newcommand{\AcpkgSyst}{\ensuremath{0.0}\xspace}

\newcommand{\Acppg}{\ensuremath{-9.4}\xspace}
\newcommand{\AcppgStat}{\ensuremath{6.6}\xspace}
\newcommand{\AcppgSyst}{\ensuremath{0.1}\xspace}

\usepackage[caption=false]{subfig}

\begin{document}

\markboth{Hector Gisbert, Marcel Golz, Dominik Stefan Mitzel}
{Theoretical and experimental status of rare charm decays}

\catchline{}{}{}{}{}

\title{THEORETICAL AND EXPERIMENTAL STATUS\\ OF RARE CHARM DECAYS}

\author{Hector Gisbert}

\address{Fakultät für Physik, TU Dortmund, Otto-Hahn-Str.\,4, D-44221 Dortmund, Germany\\
hector.gisbert@tu-dortmund.de}

\author{Marcel Golz}

\address{Fakultät für Physik, TU Dortmund, Otto-Hahn-Str.\,4, D-44221 Dortmund, Germany\\
marcel.golz@tu-dortmund.de}

\author{Dominik Stefan Mitzel}

\address{CERN, Espl. des Particules 1, CH-1211 Meyrin, Switzerland\\
dominik.mitzel@cern.ch 
}
\maketitle

\pub{Published in Mod. Phys. Lett. A \textbf{36} (2021) 2130002 }{}

\begin{abstract}
Rare charm decays offer the unique possibility to explore flavor-changing neutral-currents in the up-sector within the Standard Model and beyond. Due to the lack of effective methods to reliably describe its low energy dynamics, rare charm decays have been considered as less promising for long. However, this lack does not exclude the possibility to perform promising searches for New Physics {\it per~se}, but a different philosophy of work is required. Exact or approximate symmetries of the Standard Model allow to construct clean null-test observables, yielding an excellent road to the discovery of New Physics, complementing the existing studies in the down-sector. In this review, we summarize the theoretical and experimental status of rare charm $|\Delta c|=|\Delta u|=1$ transitions, as well as opportunities for current and future experiments such as LHCb, Belle~II, BES~III, the FCC-ee and proposed tau-charm factories. We also use the most recent experimental results to report updated limits on lepton-flavor conserving and lepton-flavor violating Wilson coefficients.

\keywords{Rare charm decays; Flavour-Changing Neutral-Currents; Flavour Physics; Physics Beyond the Standard Model.}
\end{abstract}

\footnotetext{This is an Open Access article published by World Scientific Publishing Company. It is distributed under the terms of the \href{https://creativecommons.org/licenses/by/4.0}{Creative Commons Attribution 4.0 (CC BY) License} which permits use, distribution and reproduction in any medium, provided the original work is properly cited.}

\clearpage
\section{Introduction}

Despite being consistent with an enormous amount of experimental results, there are undoubtedly phenomena that the Standard Model (SM) fails to explain and a more fundamental theory has to exist. Nowadays, the combined effort of theoreticians and experimentalists is to formulate extensions to account for the apparent shortcomings of the SM, collectively referred to as New Physics (NP) models, and to find hints for its breakdown.

Rare decays of flavored mesons containing an \squark, \cquark or \bquark quark receive contributions from flavor-changing neutral-current (FCNC) processes and are sensitive probes to heavy degrees of freedom at mass scales much higher than the available center-of-mass energies in the most powerful particle colliders. New and yet unknown particles and interactions can modify the rate of such a process, change the angular distributions of the decay products, or introduce additional sources of CP violation. Precision measurements of FCNC processes have been proven to have the potential to indirectly point towards the existence of new particles long before their direct detection. Famous examples are the first hints for the necessity of the charm quark\cite{Glashow:1970gm}, or the extremely heavy mass of the top quark, inferred from the suppression of neutral kaon decays\cite{Bott-Bodenhausen:1967vka,Foeth:1969hi}, and the observation of neutral $\Bd$ meson mixing\cite{ALBRECHT1987245}, respectively.

While in the past most of the experimental effort has focused on studying rare processes in the kaon and beauty sectors, investigations of rare charm decays, which are sensitive to $|\Delta c|=|\Delta u|=1$ transitions, have only started. Due to the low statistics available, a major part of experimental analyses has been restricted to setting upper limits on rare and forbidden decay processes. Since most decays are dominated by resonant contributions, which cannot be described in a consistent theoretical framework, rare charm decays have been considered as less promising for a long time. The presence of large uncertainties coming from effects of the strong interaction at low energies, such as hadronization and the formation of light intermediate resonances, does not prevent clean searches for NP in rare charm decays. However, a different philosophy of work is required compared to beauty and kaon physics, aiming for optimized observables.

In this review, we highlight the unique phenomenology of rare charm decays and how this can be used to search for NP in the up-sector. The SM symmetries in the charm system offer the possibility to define null-test observables with small or negligible theory uncertainties in resonance-dominated semi-leptonic and radiative decays. The possibility to investigate angular distributions, CP asymmetries and tests for lepton universality experimentally in these decays with typical branching fractions of $10^{-5}$--$10^{-7}$ has only opened recently, and precision measurements are expected to be possible in the near future at the current flavor experiments \lhcb, \belle II and \bes3 .

In addition, in light of the persistent anomalies in rare \B decays (e.g.~see Ref.~\refcite{Bifani:2018zmi} and references therein), the charm systems offers a complementary opportunity whose potential has hardly been exploited so far.

The paper is organized as follows: We start with an overview of the theoreti\-cal framework in Sec.~\ref{sec:theoFrame}, discussing the short- and long-distance description of rare charm decays. In Sec.~\ref{sec:NPmodels}, we briefly summarize the NP models that have been explored in the context of rare charm decays. We then present a summary of experimental\- searches for rare and forbidden decays in Sec.~\ref{sec:expSearches}, and how experimental limits can be translated in model-independent bounds on Wilson coefficients in Sec.~\ref{sec:modelindep}. In Sec.~\ref{sec:nulltests}, we discuss strategies to test the SM with clean null tests in semi-leptonic and radiative decays. We conclude the review with a brief outlook to future prospects in Sec.~\ref{sec:outlook} and closing remarks in Sec.~\ref{sec:conclusion}.

\section{Theoretical framework}\label{sec:theoFrame}

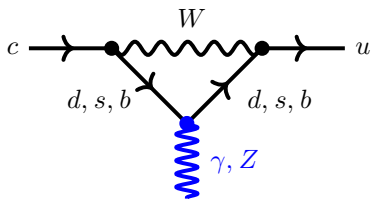
\begin{figure}[h!]
\centering

\tikzset{
    vector/.style={decorate, decoration={snake}, draw},
      gluon/.style={decorate, draw=blue,
        decoration={snake,amplitude=4pt, segment length=5pt}}, 
    fermion/.style={draw=black, postaction={decorate},
        decoration={markings,mark=at position .55 with {\arrow[draw=black]{>}}}},
}

\hskip .6cm
\begin{tikzpicture}[line width=1.5 pt, node distance=2cm and 2cm ]
	
\coordinate[label=below:\scriptsize{$ $}] (v1);
\coordinate[above=of v1,label=above:\scriptsize{$ $}] (v2);
\coordinate[right=of v2,label=above:\scriptsize{$ $}] (v3);
\coordinate[below= of v3,label=below:\scriptsize{$ $}] (v4);
\coordinate[above right= 2cm and 1.1cm of v4,label=right:$u$] (e2);
\coordinate[below right= 2cm and 1.1cm of v3,label=right:$ $] (f2);
\coordinate[above left = 2cm and 1.1cm of v1,label=left :$c$] (f1);
\coordinate[below left = 2cm and 1.1cm of v2,label=left :$ $] (e1);
\coordinate[above left=1.3cm and  -0.4cm of v1,label=left :$d\text{,}\:s\text{,}\:b$] (a1);
\coordinate[above right=0.4cm and  1.4cm of v2,label=left :$W$] (a2);
\coordinate[below right=0.7cm and  0.8cm of v3,label=left :$d\text{,}\:s\text{,}\:b$] (a3);
\coordinate[below right=0.35cm and  -0.8cm of v4,label=left :$ $] (a2);
\coordinate[below right=1.5cm and  0.1cm of v3,label=left :$\color{blue}{\gamma\text{,}\:Z}$];
\coordinate[above right=1cm and  -1cm of v4,label=left :$ $] (o1);   
\coordinate[above right=0cm and  -1cm of v4,label=left :$ $] (o2);  

		\draw[vector] (v3) -- (v2);
		\draw[fermion] (f1) -- (v2);
		\draw[fermion] (v3) -- (e2);
		\draw[fermion] (v2) -- (o1);
		\draw[fermion] (o1) -- (v3);
	    \draw[gluon] (o1) -- (o2);
		
	    \fill[black] (v2) circle (.1cm);
	    \fill[black] (v3) circle (.1cm);
	    \fill[blue] (o1) circle (.1cm);

\end{tikzpicture}
\caption{SM electroweak penguin topology contributing to $|\Delta c|=|\Delta u|=1$ transitions.}\label{fig:cu_amplitude}	
\end{figure}

In the SM, the leading contribution to $|\Delta c|=|\Delta u|=1$ transitions appears at 1-loop level, consequently its study provides an excellent window to test its quantum structure. Figure~\ref{fig:cu_amplitude} displays a possible 1-loop contribution via a $W$ boson, with internal down-type quarks (\dquark,\squark,\bquark) flowing in the loop. The amplitude of this diagram $\mathcal{A}(c\to u)$ can be written as
\begin{align}
\label{eq:ampli}
\begin{split}
    \mathcal{A}(c \to u) &\,=\,\sum_{i=d,s,b} \lambda_i\,f_i~,
\end{split}
\end{align}
where $\lambda_{i}\equiv V_{ci}^*V^{}_{ui}$ encodes the dependence on the Cabibbo--Kobayashi--Maskawa (CKM) matrix elements and thus the CP violating phenomena in the SM. The loop function $f_i\equiv f(x_i)$\,$\sim$\,$\frac{x_i}{(4\pi)^2}$ parameterizes the quantum effects with $x_i\equiv \frac{m_{i}^2}{M_W^2}$, where $m_{i}$ and ${M_W}$ are the masses of the down-type quarks and $W$ boson, respectively.

Using the unitarity condition of the CKM matrix, $\sum_{i=d,s,b}\lambda_i=0$, one can remove the dependence on $\lambda_d$, and Eq.~\eqref{eq:ampli} becomes
\begin{align}
\label{eq:gimsuppr}
\begin{split}
    \mathcal{A}(c \to u) &\,=\,\lambda_s\,\left[f_s-f_d+\xi_b\,\big(f_b-f_d\big)\right]~,
\end{split}
\end{align}
where $\xi_b\equiv\lambda_b/\lambda_s$. Equation~\eqref{eq:gimsuppr} exhibits two important features that characterize rare charm decays in the SM
\begin{itemlist}
 \item Branching fractions for rare $c\to u$ transitions are suppressed via the Glashow--Iliopoulos--Maiani (GIM) mechanism, as shown in the first term of Eq.~\eqref{eq:gimsuppr}, resulting in the na\"ve estimation of $\mathcal{A}(c\to u)\sim\mathcal{O}(10^{-8})$.
 \item CP asymmetries are CKM-suppressed through $|\xi_b|\sim 10^{-3}$, as can be seen in the second term of Eq.~\eqref{eq:gimsuppr}.
\end{itemlist}
Furthermore, the very effective GIM mechanism leads to specific angular distributions of the final states, as will become clearer later.

In summary, short-distance contributions in the SM for $|\Delta c|=|\Delta u|=1$ transitions are well below the current experimental precision due to both GIM- and CKM-suppressions. Hence, the study of observables like branching fractions, CP and angular asymmetries provides a fantastic road towards the discovery of physics beyond the SM, since large signals indicate clear signs of NP.
The above statements are based on a na\"ive estimation of the SM contribution. For the benefit of more precise calculations, a robust theoretical framework is needed. In the following sections, we give a brief summary of the current tools used to describe both the short-distance and long-distance effects involved in rare charm decays.

\subsection{Short-distance description}\label{sec:shortdist}

In the most general form, rare $|\Delta c|=|\Delta u|=1$ transitions are described by the following effective Hamiltonian\cite{Greub:1996wn,Fajfer:2002gp,deBoer:2016dcg}{:} 
\begin{align}\label{eq:Heff}
\mathcal{H}_{\rm eff} = -\frac{4\,G_F}{\sqrt2} \frac{\alpha_e}{4\pi} &\left[\sum_{i\neq T,T5} \biggl( {\cal C}_i(\mu)\, {\cal O}_i(\mu) + {\cal C}_i^\prime(\mu)\, {\cal O}_i^\prime(\mu) \biggr) + \sum_{i=T,T5} {\cal C}_i(\mu) \,{\cal O}_i(\mu) \right]\,,
\end{align}
with the following local dimension-six operators\cite{Chetyrkin:1996vx,Bobeth:1999mk,Gambino:2003zm}{:}
\begin{equation}
\begin{split}
{\cal O}_7 &= {m_c \over e} (\ubar_L \sigma_{\mu\nu} c_R) F^{\mu\nu} \,, \\
{\cal O}_9 &= (\ubar_L \gamma_\mu c_L) (\lbar \gamma^\mu \ell) \,, \\ 
{\cal O}_{S\,(P)} &= (\ubar_L c_R) (\lbar(\gamma_5)\ell) \,,
\end{split}
\begin{split}
{\cal O}_8 &= {m_c \,g_s \over e^2} (\ubar_L \sigma_{\mu\nu}\,T^a c_R)\, G^{\mu\nu}_a \,, \\
{\cal O}_{10} &= (\ubar_L \gamma_\mu c_L) (\lbar \gamma^\mu \gamma_5 \ell) \,,  \\
{\cal O}_{T\,(T5)} &= {\textstyle \frac{1}{2}} (\ubar \sigma_{\mu\nu} c) (\lbar \sigma^{\mu\nu} (\gamma_5)\ell) \,,
\end{split}
\label{eq:operators}
\end{equation}
where $q_{L,R}=\frac{1}{2}(1\mp\gamma_5)q$ are chiral quark fields, $T^a$ are the generators of SU$(3)_C$ and $g_s$ is the strong coupling. Furthermore, $G_F$ is the Fermi constant and $\alpha_e=e^2/(4\pi)$ stands for the fine-structure constant with the electromagnetic coupling $e$. Finally, $\sigma^{\mu\nu}=\frac{i}{2}[\gamma^\mu,\gamma^\nu]$ and $F^{\mu\nu},G_a^{\mu\nu}$ with $a=1,\ldots,8$ denote the electromagnetic and gluconic field strength tensor, respectively. Primed operators are obtained replacing $L(R) \to R(L)$ in Eq.~\eqref{eq:operators}. Note that Eq.~\eqref{eq:Heff} is a direct consequence of the operator product expansion (OPE), which allows the factorization between the matrix elements of local operators $\mathcal{O}_i$ and the Wilson coefficients $\mathcal{C}_i$, which parameterize the strength of each operator. The operators given by Eq.~\eqref{eq:operators} are constructed with the light fields with masses below $\mu<m_b$, with $\mu$ being the renormalization scale. The heavy fields with masses greater than $\mu \gtrsim m_b$ are removed as dynamical degrees of freedom. Effects of the heavy fields are implicitly encoded in the Wilson coefficients. Hence, experimental deviations of these coefficients from their SM predictions indicate a signal of physics beyond the SM (BSM), since such a discrepancy requires additional dynamical degrees of freedom at high-energy scales.

The general OPE setup ({i.e.} definition of the operator basis, matching of the SM contributions onto the effective theory, renormalization group (RG) evolution of Wilson coefficients from the high-energy to the low-energy scale) needed for the computation of the Wilson coefficients relevant in $|\Delta c|=|\Delta u|=1$ transitions is almost analogous to $|\Delta b|=|\Delta s|=1$ transitions\cite{Buchalla:1995vs,Borzumati:1998tg,Chetyrkin:1996vx,Chetyrkin:1997gb,Bobeth:1999mk,
Gambino:2003zm,Gorbahn:2004my,Gorbahn:2005sa}. However, due to the specific CKM and mass structure of charm FCNCs, the predictions for this sector can differ from each other by several orders of magnitude depending on which corrections are taken into account. Reference~\refcite{deBoer:2016dcg} provides a consistent expansion of the complete SM computation for rare charm transitions to $\mathcal{O}(\alpha_s)$ with $\alpha_s=g_s^2/(4\pi)$. In the following, we give an overview of the steps required to achieve a perturbative next-to-(next-to-) leading order precision of the short-distance contributions\cite{deBoer:2016dcg}. These are:
\begin{enumerate}[label=(\roman*)]
    \item matching of SM contributions onto Weak Effective Theory at $\mu=M_W$, 
    \item RG-evolution of Wilson coefficients from $M_W$ to $m_b$,
    \item integrating out the $b$ quark and second matching 
    at $\mu=m_b$,
    \item RG-evolution of Wilson coefficients from $m_b$ to the charm scale $\mu_c$.
\end{enumerate}

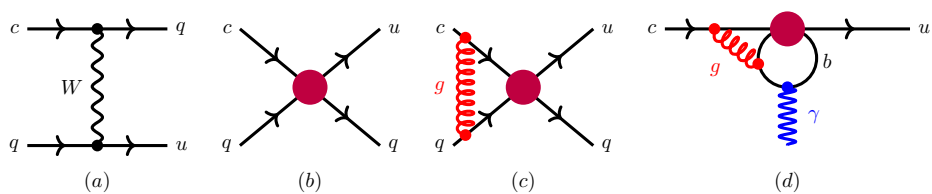
\begin{figure}[h!]
\centering
  \resizebox{\textwidth}{!}{  
\subfloat{\label{fig:cu_amplitude:a}}{\tikzset{
    vector/.style={decorate, decoration={snake}, draw},
      gluon/.style={decorate, draw=blue,
        decoration={snake,amplitude=4pt, segment length=5pt}}, 
    fermion/.style={draw=black, postaction={decorate},
        decoration={markings,mark=at position .55 with {\arrow[draw=black]{>}}}},
}

\begin{tikzpicture}[line width=1.5 pt, node distance=2cm and 2cm ]
	
\coordinate[label=below:\scriptsize{$ $}] (v1);
\coordinate[above=of v1,label=above:\scriptsize{$ $}] (v2);
\coordinate[right=of v2,label=above:\scriptsize{$ $}] (v3);
\coordinate[below= of v3,label=below:\scriptsize{$ $}] (v4);
\coordinate[above right= 2cm and .2cm of v4,label=right:$q$] (e2);
\coordinate[below right= 2cm and .2cm of v3,label=right:$u$] (f2);
\coordinate[above left = 2cm and .2cm of v1,label=left :$c$] (f1);
\coordinate[below left = 2cm and .2cm of v2,label=left :$q$] (e1);
\coordinate[above left=1cm and  -0.9cm of v1,label=left :$W$] (a1);
\coordinate[below right=0.35cm and  -0.8cm of v4,label=left :$ $] (a2);
\coordinate[above right=2cm and  -1cm of v4,label=left :$ $] (o1);   
\coordinate[above right=0cm and  -1cm of v4,label=left :$ $] (o2);  
\coordinate[below=  0.3 cm of o2,label=below:{$(a)$}] (caption);

		\draw[fermion] (e1) -- (o2);
		\draw[fermion] (f1) -- (o1);
		\draw[fermion] (o1) -- (e2);
		\draw[fermion] (o2) -- (f2);
	    \draw[vector] (o1) -- (o2);
	    \fill[black] (o1) circle (.1cm);
	    \fill[black] (o2) circle (.1cm);
	    
\end{tikzpicture}}

\subfloat{\label{fig:cu_amplitude:b}}{
  
  \tikzset{
    vector/.style={decorate, decoration={snake}, draw},
      gluon1/.style={decorate, draw=blue,
        decoration={snake}, draw},
    fermion/.style={draw=black, postaction={decorate},
        decoration={markings,mark=at position .55 with {\arrow[draw=black]{>}}}},
                jon/.style={draw=Green, postaction={decorate}},
                gluon/.style={decorate, draw=red,
        decoration={coil,amplitude=4pt, segment length=5pt}}, 
}

\begin{tikzpicture}[line width=1.5 pt, node distance=2cm and 2cm ]
	
\coordinate[label=below:\scriptsize{$ $}] (v1);
\coordinate[above=of v1,label=above:\scriptsize{$ $}] (v2);
\coordinate[right=of v2,label=above:\scriptsize{$ $}] (v3);
\coordinate[below= of v3,label=below:\scriptsize{$ $}] (v4);
\coordinate[above right= 2cm and .2cm of v4,label=right:$u$] (e2);
\coordinate[below right= 2cm and .2cm of v3,label=right:$q$] (f2);
\coordinate[above left = 2cm and .2cm of v1,label=left :$c$] (f1);
\coordinate[below left = 2cm and .2cm of v2,label=left :$q$] (e1);

\coordinate[below right=0.35cm and  -0.8cm of v4,label=left :$ $] (a2);
\coordinate[above left=1cm and  -1cm of v1,label=left:$ $] (o1);   
\coordinate[above right=0cm and  -1cm of v4,label=left :$ $] (o2);  
\coordinate[above left=1cm and  0.2cm of v1,label=left :$ $] (a1);
\coordinate[below=  0.3 cm of o2,label=below:{$(b)$}] (caption);

		\draw[fermion] (e1) -- (o1);
		\draw[fermion] (f1) -- (o1);
		\draw[fermion] (o1) -- (e2);
		\draw[fermion] (o1) -- (f2);
	    \fill[purple] (o1) circle (.3cm);
	    
\end{tikzpicture}}

\subfloat{\label{fig:cu_amplitude:c}}{
  
  \tikzset{
    vector/.style={decorate, decoration={snake}, draw},
      gluon1/.style={decorate, draw=blue,
        decoration={snake}, draw},
    fermion/.style={draw=black, postaction={decorate},
        decoration={markings,mark=at position .55 with {\arrow[draw=black]{>}}}},
                jon/.style={draw=Green, postaction={decorate}},
                gluon/.style={decorate, draw=red,
        decoration={coil,amplitude=4pt, segment length=5pt}}, 
}

\begin{tikzpicture}[line width=1.5 pt, node distance=2cm and 2cm ]
	
\coordinate[label=below:\scriptsize{$ $}] (v1);
\coordinate[above=of v1,label=above:\scriptsize{$ $}] (v2);
\coordinate[right=of v2,label=above:\scriptsize{$ $}] (v3);
\coordinate[below= of v3,label=below:\scriptsize{$ $}] (v4);
\coordinate[above right= 2cm and .2cm of v4,label=right:$u$] (e2);
\coordinate[below right= 2cm and .2cm of v3,label=right:$q$] (f2);
\coordinate[above left = 2cm and .2cm of v1,label=left :$c$] (f1);
\coordinate[below left = 2cm and .2cm of v2,label=left :$q$] (e1);
\coordinate[below right=0.35cm and  -0.8cm of v4,label=left :$ $] (a2);
\coordinate[above left=1cm and  -1cm of v1,label=left:$ $] (o1);   
\coordinate[above right=0cm and  -1cm of v4,label=left :$ $] (o2);  
\coordinate[above left=1cm and  0.2cm of v1,label=left :$\color{red}{g}$] (a1);
\coordinate[below=  0.3 cm of o2,label=below:{$(c)$}] (caption);

		\draw[fermion] (e1) -- (o1);
		\draw[fermion] (f1) -- (o1);
		\draw[fermion] (o1) -- (e2);
		\draw[fermion] (o1) -- (f2);
		\draw[gluon] (0.01,0.18) -- (0.01,1.85);
	    \fill[red] (0.01,0.18) circle (.1cm);
	    \fill[red] (0.01,1.85) circle (.1cm);
	    \fill[purple] (o1) circle (.3cm);
	    
\end{tikzpicture}}

\subfloat{\label{fig:cu_amplitude:d}}{
	
\tikzset{
    vector/.style={decorate, decoration={snake}, draw},
      gluon1/.style={decorate, draw=blue,
        decoration={snake,amplitude=4pt, segment length=5pt}}, 
    fermion/.style={draw=black, postaction={decorate},
        decoration={markings,mark=at position .55 with {\arrow[draw=black]{>}}}},
    fermion1/.style={draw=black, postaction={decorate},
        decoration={markings,mark=at position .25 with {\arrow[draw=black]{>}}}},
    gluon/.style={decorate, draw=red,
        decoration={coil,amplitude=4pt, segment length=5pt}},
}

\begin{tikzpicture}[line width=1.5 pt, node distance=2cm and 2cm ]
	
\coordinate[label=below:\scriptsize{$ $}] (v1);
\coordinate[above=of v1,label=above:\scriptsize{$ $}] (v2);
\coordinate[right=of v2,label=above:\scriptsize{$ $}] (v3);
\coordinate[below= of v3,label=below:\scriptsize{$ $}] (v4);
\coordinate[above right= 2cm and 1.1cm of v4,label=right:$u$] (e2);
\coordinate[below right= 2cm and 1.1cm of v3,label=right:$ $] (f2);
\coordinate[above left = 2cm and 1.1cm of v1,label=left :$c$] (f1);
\coordinate[above left=1.5cm and  -0.5cm of v1,label=left :$ $] (a1);
\coordinate[above right=0.0cm and  1.0cm of v2] (a2);
\coordinate[below right=0.5cm and  1.0cm of v2] (a0);
\coordinate[below right=0.55cm and  -0.1cm of v3,label=left :$b$] (a3);
\coordinate[below right=1.5cm and  -0.3cm of v3,label=left :$\color{blue}{\gamma}$];
\coordinate[above right=1cm and  -1cm of v4,label=left :$ $] (o1);   
\coordinate[above right=0cm and  -1cm of v4,label=left :$ $] (o2);  
\coordinate[below=  0.3 cm of o2,label=below:{$(d)$}] (caption);

\coordinate[above left=1.3cm and  0.0cm of v1,label=left :$\color{red}{g}$];

		\draw[fermion1] (f1) -- (a2);
		\draw[fermion] (a2) -- (e2);
	    \draw[gluon1] (o1) -- (o2);

	    \draw[black] (a0) circle (.5cm);	    
	    \fill[purple] (a2) circle (.3cm);
	    
	    \draw[gluon] (0.5,1.4) -- (-0.25,2);
	    
	    \fill[red] (0.5,1.4) circle (.1cm);
	    \fill[red] (-0.25,2) circle (.1cm);

	    \fill[blue] (o1) circle (.1cm);

\end{tikzpicture}}

}

\caption{Exemplary contributions to $|\Delta c|=|\Delta u|=1$ transitions at different energy scales. Current--current topology shown in (a). Current--current topologies when the $W$ boson is integrated out, with and without gluonic (denoted by $g$) corrections (c) and (b), respectively. Photon $\gamma$ penguin contribution generated at the $b$ quark mass threshold shown in (d).}	\label{fig:cu_amplitude_all}
\end{figure}

In the SM, $|\Delta c|=|\Delta u|=1$ transitions are driven via the exchange of a $W$ boson between two weak charged quark currents as shown in Fig.~\ref{fig:cu_amplitude_all}(a). At high energies, where quantum chromodynamics (QCD) corrections are small, however, still below the scale of electroweak symmetry breaking, the $W$ boson can be integrated out, and the interaction can be well described in terms of a single operator
\begin{align}\label{eq:current1}
    \begin{split}
    {\cal O}_2^q\,&=\,(\bar{u}_L\,\gamma_\mu\,q_L)\,(\bar{q}_L\,\gamma^\mu\,c_L)~,
    \end{split}
\end{align}
as shown Fig.~\ref{fig:cu_amplitude_all} $(b)$. 

With decreasing energies, $\mu<M_W$, the gluonic corrections become sizeable and have to be summed up using the OPE and the renormalization group equations (RGEs).\cite{Gorbahn:2004my,Czakon:2006ss,Gorbahn:2005sa} This procedure requires an additional operator, see Fig.~\ref{fig:cu_amplitude_all}(c), which mixes under renormalization with ${\cal O}_2^q$,
\begin{align}\label{eq:current2}
    \begin{split}
    {\cal O}_1^q\,&=\,(\bar{u}_L\,\gamma_\mu\,T^a\,q_L)\,(\bar{q}_L\,\gamma^\mu\,T^a\,c_L)~.
    \end{split}
\end{align}
The operators \eqref{eq:current1} and \eqref{eq:current2} are called current--current operators and are also responsible for non-leptonic $D$-meson decays.

At $\mu>m_b$, the treatment of light quarks as massless leads to a fully effective GIM mechanism causing the cancellation of penguin contributions\cite{Greub:1996wn}, as already seen in Eq.~\eqref{eq:gimsuppr}. Penguin contributions appear when the $b$ quark is integrated out as an effective degree of freedom at its threshold, see Fig.~\ref{fig:cu_amplitude_all}(d). After the matching, the Wilson coefficients are evolved from $m_b$ to $m_c$ using the RGEs, which induces mixing between different $|\Delta c|=|\Delta u|=1$ operators, given by Eq.~\eqref{eq:operators}.\cite{deBoer:2015boa}

The last step is the evaluation of the Wilson coefficients down to $m_c$. The computation of the matrix elements associated to certain operators result in ``effective'' coefficients which account for effects of nonzero light quark masses. Strictly speaking, the effective coefficients are not Wilson coefficients, since they contain low-energy effects. Wilson coefficients are process universal, independent of the operator basis, regularization and renormalization schemes\cite{Misiak:1992bc}. The effective coefficients $\mathcal{C}_{7,9}^{\rm eff}$, which are the only nonvanishing ones, at the charm scale $\mu_c$ read\cite{deBoer:2015boa,Buras:1994dj}
\begin{equation}
\begin{split}
\mathcal{C}_7^{\,\rm eff}(q^2 \approx 0) &\simeq -0.0011-0.0041\,\text{i} \,, \\
\mathcal{C}_9^{\,\rm eff} (q^2) &\simeq -0.021 \bigg[V_{cd}^*V_{ud}\,L(q^2,m_d,\mu_c) + V_{cs}^*V_{us}\,L(q^2,m_s,\mu_c)\bigg] \,,
\end{split}
\label{eq:CSM}
\end{equation}
where $q^2$ is the dilepton invariant mass squared and $L(q^2,m_q,\mu_c)$ is a function that accounts for the low dynamical effects due to $m_q\neq 0$. The explicit form of $L(q^2,m_q,\mu_c)$ can be found in Ref.~\refcite{deBoer:2015boa}. The coefficients $\mathcal{C}_{7,9}^{\rm eff}$ are dominated by the matrix elements $\mathcal{O}_{1,2}^q$. The $q^2$-dependence of $\mathcal{C}_7^{\rm eff}(q^2\approx 0)$ is negligible. At $\mu_c=m_c$, one obtains $|\mathcal{C}_9^{\rm eff}|\lesssim 0.01$, while
$\text{Im}[\mathcal{C}_7^{\rm eff}]$ increases from $-0.004$ at $q^2=0$ to $-0.001$ at high $q^2$\cite{deBoer:thesis}.

All other SM Wilson coefficients in Eq.~\eqref{eq:Heff} do not receive any contribution,\footnote{QED corrections contribute to Eq.~\eqref{eq:WCsSMzero}, however they are negligible with respect to the results from QCD, i.e.~$\mathcal{C}_{10}^{\text{QED}}(\mu_c)<0.01\,\mathcal{C}_{9}(\mu_c)$, see Ref.~\refcite{deBoer:thesis}.}
\begin{align}\label{eq:WCsSMzero}
   \mathcal{C}_i^{\prime\,\text{SM}}=\mathcal{C}_{S}^{\text{SM}}=\mathcal{C}_T^{\text{SM}}=\mathcal{C}_{T5}^{\text{SM}}=\mathcal{C}_{10}^{\text{SM}}=0~. 
\end{align}
Especially $\mathcal{C}_{10}^{\text{SM}}=0$ distincts charm FCNCs from $K$ or $B$ physics. Since $\mathcal{C}_{10}$ corresponds to an axial vector coupling for the leptonic part, effects on the $V$--$A$ structure of the SM are shut off at the charm scale and charm physics within the SM is dominated by QCD and quantum electrodynamics (QED) effects.

\subsection{Long-distance description}\label{sec:longdist}

The OPE allows for the separation of short-distance and long-distance effects. In order to fully assess the decay amplitude of rare $c\to u \ell^+ \ell^-$ transitions, besides the Wilson coefficients ${\cal C}_i(\mu)$, we also need to determine the hadronic matrix elements $\langle{\cal O}_i(\mu)\rangle$ which encode the non-perturbative dynamics at low energies. Since measurable observables cannot depend on the renormalization scale, the dependence on $\mu$ has to cancel in the product of ${\cal C}_i(\mu)$ and $\langle{\cal O}_i(\mu)\rangle$. A proper understanding of the long-distance dynamics with a solid effective field theory framework is still missing in the literature, mainly due to $\Lambda_{\rm QCD}\sim m_c$, which makes a perturbative expansion in powers of $1/m_c$ slowly converging at best. In the following, we summarize the available techniques to deal with this challenging task.

\subsubsection{Form factors}\label{sec:FFs}

\begin{table}[t]
\tbl{Number of form factors (FFs) or decay constant for different modes as well as their availability from different sources. $^\dagger$ In the limit of $m_\ell=0$.}{
\resizebox{\textwidth}{!}{ 
\begin{tabular}{@{}ccccc@{}}
\toprule
Mode & $\#$ of FFs  & Reference \\
\colrule
$D^0\to\ell^+\ell^-$ &  decay constant & Lattice\cite{Aoki:2019cca}$\,|\,|\,$ Exp.\cite{Amhis:2019ckw}\\
$D^+\to\pi^+\,\ell^+\ell^-$ &  3 & Lattice \cite{Lubicz:2017syv,Lubicz:2018rfs}$\,|\,|\,$ Exp.\cite{Amhis:2019ckw} \\
$D^0\to\pi^0\,\ell^+\ell^-$ &  3 & Lattice \cite{Lubicz:2017syv,Lubicz:2018rfs}$\,|\,|\,$ Exp.\cite{Amhis:2019ckw} \\
$D^+_s\to K^+\,\ell^+\ell^-$ &  3 & Lattice\cite{Lubicz:2017syv,Lubicz:2018rfs,Koponen:2012di}$\,|\,|\,$ Exp.\cite{Amhis:2019ckw} \\
$D\to P_1 P_2 \,\ell^+\ell^-$ & 7$\,{^\dagger}$ &  HH$\chi$PT\cite{Das:2014sra,deBoer:2018buv,Lee:1992ih} $\,\&\,$ Exp. input\cite{Lees:2013zna}\\
$D\to\rho \,\ell^+\ell^-$ & 7$\,{^\dagger}$   &  Lattice\cite{Flynn:1997ca,Gill:2001jp} $\,|\,|\,$ LCSR\cite{Wu:2006rd} $\,|\,|\,$ CLFQM\cite{Verma:2011yw}$\,|\,|\,$ CQM\cite{Melikhov:2000yu} $\,|\,|\,$ Exp.\cite{CLEO:2011ab}  \\
$D\to\omega \,\ell^+\ell^-$ &  7$\,{^\dagger}$  &   LCSR\cite{Wu:2006rd}$\,|\,|\,$ CLFQM\cite{Verma:2011yw}$\,|\,|\,$ CQM\cite{Melikhov:2000yu} $\,|\,|\,$ Exp.\cite{Ablikim:2015gyp}   \\
$D_s\to K^* \,\ell^+\ell^-$ &  7$\,{^\dagger}$ &  LCSR\cite{Wu:2006rd} $\,|\,|\,$ CLFQM\cite{Verma:2011yw}$\,|\,|\,$ CQM\cite{Melikhov:2000yu} $\,|\,|\,$ Spectator invariance\cite{deBoer:thesis,Ablikim:2015mjo} \\
$\Lambda_c\to p\,\ell^+\ell^-$ &  10$\,{^\dagger}$ & Lattice\cite{Meinel:2017ggx} \\
\botrule
\end{tabular}\label{tab:FFs} }}
\end{table}

Form factors (FFs) parameterize our lack of knowledge about the hadronic effects in a hadronic transition. The hadronization of the $|\Delta c|=|\Delta u|=1$ operators given by Eq.~\eqref{eq:operators} leads to the factorization between the lepton and quark currents,
\begin{align}
    \langle h_c|\mathcal{O}_i|F\,\ell^+\ell^-\rangle=\langle h_c| H_{\alpha_1,...,\alpha_n}^i|F\rangle\,\langle 0| L^{\alpha_1,...,\alpha_n}_i|\ell^+\ell^-\rangle~,
\end{align}
where $h_c$ represents a charmed hadron\footnote{The inclusion of charge-conjugate decays is implied unless stated differently.} and $F$ is the final state. $H_{\alpha_1,\ldots,\alpha_n}^i$ and $L^{\alpha_1,\ldots,\alpha_n}_i$ are the quark and lepton currents of $\mathcal{O}_i$, respectively, $\alpha_1,\ldots,\alpha_n$ being shared Lorentz indices between both currents. The quantity $\langle 0| L^{\alpha_1,\ldots,\alpha_n}_i|\ell^+\ell^-\rangle$ can be computed applying perturbation theory in QED, while effects contained in $\langle h_c| H_{\alpha_1,\ldots,\alpha_n}^i|F\rangle$ require non-perturbative techniques to deal with the dynamics of QCD at low energies. Even though there is no rigorous effective field theory to face these effects, we can reduce the problem using symmetries at low energies. In particular, requiring Lorentz structure and parity invariance of QCD, the vectorial~V $(n=1)$ current in which a $D$ meson decays into a pseudoscalar $P$ can be written as
\begin{align}
    \langle D(p_D)|\ubar\, \gamma_\alpha\, c|P(p_P)\rangle\,=\,p_\alpha\,F_+^{\rm V}(q^2) +q_\alpha\,F_-^{\rm V}(q^2)~,
\end{align}
where the functions $F_\pm^{\rm V}$ are the FFs associated to the $\rm V$ current, and the four-momenta $q,\,p$ are given by $q^\alpha=(p_D-p_P)^\alpha$ and $p^\alpha=(p_D+p_P)^\alpha$, respectively. Due to four-momentum conservation, FFs can only depend on the four-momentum transfer $q$. In contrast, for a tensorial $\rm T$ ($n=2$) current, due to its antisymmetric Lorentz structure, only one FF appears
\begin{align}
    \langle D(p)|\ubar\, \sigma_{\alpha\beta}\, c|P(k)\rangle\,=\,\text{i}\,\left(p_\alpha\,q_\beta-q_\alpha p_\beta\right)F^{\rm T}(q^2)~.
\end{align}
\noindent
Currently, FFs can only be computed using methods such as lattice gauge theory, QCD light-cone sum rules (LCSR)\footnote{LCSR calculations are valid only at low $q^2$.} or fitting theory/models\footnote{For instance, the heavy hadron chiral perturbation theory (HH$\chi$PT)\cite{Ablikim:2015mjo}, the covariant light front quark model (CLFQM)\cite{Verma:2011yw} and the constituent quark model (CQM)\cite{Melikhov:2000yu}.} to experimental data. Table~\ref{tab:FFs} summarizes the number of FFs (second column) needed for exemplary decays modes (first column), as well as how they can be extracted (third column).\footnote{Using spectator invariance one can relate the FFs of $D_s\to K^*$ to $D \to (\rho,\omega)$ FFs.\cite{deBoer:thesis,Ablikim:2015mjo}}

\subsubsection{Resonance contributions} \label{sec:resoContributions}

Often charm decay modes are dominated by resonance contributions. These effects can be parameterized by fitting Breit--Wigner distributions to experimental data. The main contribution in $h_c\to F\ell^+\ell^-$ decays results from $h_c\to FM^*$ with $M^*\to \gamma^* \to \ell^+\ell^-$ which induces resonance effects on $\mathcal{O}_{9,P}$ that can be phenomenologically parameterized by\cite{deBoer:2015boa,Fajfer:2005ke}
\begin{align}
\mathcal{C}_9^R &= a_\rhoz \,e^{\text{i}\,\delta_\rhoz} \biggl( {1 \over q^2 - m_\rhoz^2 + \text{i}\,m_\rhoz\Gamma_\rhoz} - {1\over3} {1 \over q^2 - m_\omega^2 + \text{i}\,m_\omega\Gamma_\omega} \biggr) + {a_\phi \,e^{\text{i}\,\delta_\phi} \over q^2 - m_\phi^2 + \text{i}\,m_\phi\Gamma_\phi} \,, \nonumber\\
\mathcal{C}_P^R &= {a_\eta \,e^{\text{i}\,\delta_\eta} \over q^2 - m_\eta^2 + \text{i}\,m_\eta\Gamma_\eta} +{a_{\eta^\prime} \over q^2 - m_{\eta^\prime}^2 + \text{i}\,m_{\eta^\prime}\Gamma_{\eta^\prime}} \,,
\label{eq:C9_CP_res}
\end{align}
where $a_M$ is the resonance parameter with $M=\rho^0,\phi,\eta,\eta^\prime,$\footnote{Along this work, we use the following abbreviations for the resonances: $\rhoz(770) \equiv \rhoz$, $\omegaz(782) \equiv \omegaz$, $\Kstarz(892) \equiv \Kstarz$, $\phi(1020) \equiv \phi$, $\Kstarp(892) \equiv \Kstarp$, $\rhop(770) \equiv \rhop$, $\etapr(958) \equiv \etapr$.} and isospin has been employed in Eq.~\eqref{eq:C9_CP_res} to relate the $\rho^0$ to the $\omega$. Here, $m_M$ and $\Gamma_M$ denote the mass and the total decay rate of $M$, respectively. The $a_M$ parameters can be extracted from measurements of branching fractions $\mathcal{B}(h_c\to F M)$ and $\mathcal{B}(M \to \ell^+ \ell^-)$, and are given in Table~\ref{tab:a_res} for some examples studied recently. The strong phases $\delta_{\rhoz,\phi,\eta}$ are the largest source of uncertainty. They can only be constrained if experimental input on branching fractions in resonance-dominated regions is provided, which is often not the case. In most numerical analyses, they are varied between $-\pi$ and $\pi$.

The theoretical description of resonance-dominated decays is challenging and requires further investigations. Studies of these effects in the context of QCD factorization can be found in Refs.~\refcite{Feldmann:2017izn,Beylich:2011aq,Bharucha:2020eup}.

\begin{table}[t]
\tbl{Phenomenological resonance parameters (in GeV$^2$) extracted from measurements of $\mathcal{B}(h_c\to F M)$ with resonances $M=\rhoz,\phi,\eta,\eta^\prime$ decaying to $\mu^+\mu^-$. In the notation $h_c\to F\,[\mumu]_M$, the two muons arise from the intermediate resonance $M$. Details can be found in Refs.~\protect\refcite{deBoer:2018buv,Meinel:2017ggx,Bause:2019vpr,Landsberg:1986fd}.}{
\label{tab:a_res}
\centering
\resizebox{\textwidth}{!}{ 
\begin{tabular}{lcccccccc} 
\toprule
Mode && $a_\rho$ && $a_\phi$ && $a_\eta$ && $a_{\eta^\prime}$ \\
\hline
$D^+\to\pi^+$  [$\mumu]_M$  && $0.18\pm0.02$ && $0.23\pm0.01$ && $(5.7\pm0.4)\times10^{-4}$ && $\sim8\times10^{-4}$ \\
$D^0\to \pi^0$ [$\mumu]_M$&& $0.86\pm 0.04$ && $0.25\pm 0.01$ && $(5.3\pm0.4)\times10^{-4}$ && $\sim8\times10^{-4}$ \\
$D_s^+\to K^+$[$\mumu]_M$ && $0.48\pm 0.04$ && $0.07\pm 0.01$ && $(5.9 \pm 0.7)\times10^{-4}$ && $\sim7\times10^{-4}$ \\
$D^0\to \pi^+\,\pi^-$ [$\mumu]_M$ &&  $\sim0.7$&& $\sim0.3$  && $\sim0.001$ && $\sim0.001$ \\
$D^0\to K^+\,K^-$ [$\mumu]_M$ && $\sim0.5$ && $\sim0.0$ && $\sim 3\times 10^{-4}$ && --  \\
$\Lambda_c\to p$ [$\mumu]_M$ && $0.20 \pm 0.04 $  && $0.111\pm0.008$ && -- && --\\
\botrule
\end{tabular}}}
\end{table}
Resonance contributions in radiative decays $D\to V\gamma$, with $V$ a vector meson, can be included through phenomenological models. For example, the model considered in Refs.~\refcite{Fajfer:1997bh} and \refcite{Fajfer:1998dv} is a mixture of factorization, heavy quark effective theory and chiral theory. The parameters of this model are unknown and have to be extracted from experimental data, allowing implicitly for a breaking of the SU$(3)_F$ flavor symmetry.

\begin{figure}[t]
\centering
\includegraphics[width=1.\textwidth]{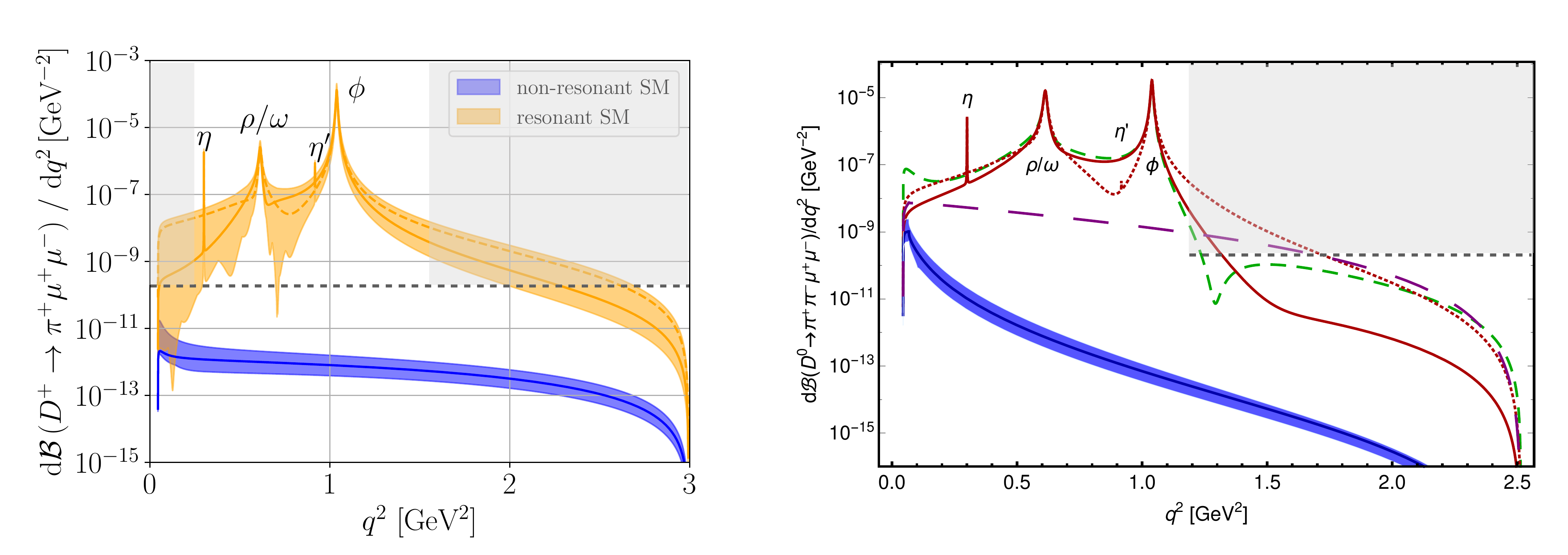}
\caption{Differential SM branching fractions of d$\,\mathcal{B}(D^+\to \pi^+ \mu^+\mu^-)/\text{d}q^2$ (left) and d$\,\mathcal{B}(D^0\to \pi^+\pi^- \mu^+\mu^-)/\text{d}q^2$ (right). The blue bands represent the non-resonant contributions including theoretical uncertainties of hadronic form factors at the charm scale $\mu_c$. 
The resonant contribution is displayed in orange (left) including the full uncertainties from form factors, resonance parameters and their associated strong phases. The solid and dashed lines (right) represent different set-ups of the strong phases $\delta_\rho$ and $\delta_\phi$. The grey dashed line shows the approximate experimental upper limits on the branching fractions taken from Refs.~\protect\refcite{Aaij:2020wyk,Aaij:2017iyr}, which are determined using restricted ranges in dimuon mass (grey shaded areas) by vetoing resonance-dominated regions. In the left, the limit has been extrapolated to the full range. Find details in Sect.~\ref{sec:SLdecays} and~\ref{sec:RasonantDecays}. The plots are adapted from Refs.~\protect\refcite{deBoer:2018buv,Bause:2019vpr}.}\label{fig:brlongdist}
\end{figure}

For instance, Fig.~\ref{fig:brlongdist} displays the differential branching fractions of
${\rm d}\mathcal{B}(D^+\to \pi^+ \mu^+\mu^-)/{\rm d}q^2$ (left) and
${\rm d}\mathcal{B}(D^0\to \pi^+\pi^- \mu^+\mu^-)/{\rm d}q^2$ (right) in the SM. The non-resonant contributions (in blue) are orders of magnitude below the resonance contributions\-, which makes them non-accessible for experiments. However, current experimental bounds still allow for large NP effects at large $q^2$ for most decay modes.

Since sensitivities of current experimental searches (gray shaded areas) are close to the orange/red resonant curves, searching for NP in branching fractions is\break a challenging endeavor, as interference effects between NP and long-distance contributions have to be taken into account and increase the theoretical uncertainties in the interpretation of NP contributions. Despite the challenges of discovering NP in branching fractions, upper limits can still serve to provide constraints on NP, as we will see in Sec.~\ref{sec:modelindep}. Furthermore, branching fraction measurements of resonance-dominated regions give insight into QCD at low energies and help to constrain strong phases, for instance.

%

\section{New Physics models generating rare charm decays} \label{sec:NPmodels}

New particles and interactions from various NP models are suitable to generate $|\Delta c|=|\Delta u|=1$ transitions at tree- or loop-level. Before discussing the experimental status of rare charm decays, in this section, we briefly summarize the literature on model-dependent BSM studies. Examples are models with non-minimal Higgs sector, such as two Higgs doublet models\cite{Ivanov:2017dad}, see Refs.~\refcite{Fajfer:2001sa,Fajfer:2015mia,Guo:2017edd}, or little Higgs \hbox{models}\cite{ArkaniHamed:2002qy,Han:2003wu,Schmaltz:2002wx}, see Refs.~\refcite{Fajfer:2005ke} and \refcite{Paul:2011ar,Fajfer:2006yc,Bigi:2011em}. Further studies are available in the context of super-symmetric models\cite{Martin:1997ns}, see Refs.~\refcite{Fajfer:2002gp,Bause:2019vpr,Fajfer:2001sa} and \refcite{Burdman:2001tf,Wang:2014uiz,Fajfer:2007dy,Cappiello:2012vg,Prelovsek:2000xy,deBoer:2018zhz} and
extra dimensions\cite{Randall:1999ee}, see Refs.~\refcite{Delaunay:2012cz} and \refcite{Paul:2012ab}. In the past, additional work has been devoted to unparticle\cite{Georgi:2007si,Luo:2007bq,Georgi:2007ek},
additional up vector-like quark singlet and models with weak vector triplets, where details can be found in Refs.~\refcite{Fajfer:2015mia,Guo:2017edd} and \refcite{Fajfer:2007dy}.
Special interest has been triggered recently by several extensions of the SM including leptoquarks\cite{Buchmuller:1986zs,Davidson:1993qk} or nonuniversal $Z^\prime$ mediators\cite{London:1986dk}, which are viable candidates to explain the flavor anomalies seen in the beauty sector\cite{Hiller:2016kry,Bifani:2018zmi}. Studies related to leptoquarks can be found in Refs.~\refcite{Bharucha:2020eup,deBoer:2015boa,deBoer:2018buv,Bause:2019vpr,deBoer:2018zhz} and \refcite{Fajfer:2008tm,Sahoo:2017lzi,Bause:2020xzj,deBoer:2017que}, those investigating models comprising $Z^\prime$ candidates are available in Refs.~\refcite{deBoer:2018buv,Bause:2019vpr,Fajfer:2015mia,Guo:2017edd,Cappiello:2012vg,Sahoo:2017lzi} and \refcite{Bause:2020xzj}. In this review, we mainly focus on a model-independent description of rare charm decays.

\section{Experimental searches for rare and forbidden decay modes} \label{sec:expSearches}

Searches for rare and forbidden decays have been carried out investigating purely leptonic final states, semi-leptonic multi-body decays of charged (\Dp, \Ds) and neutral (\Dz) mesons, as well as decays of charmed baryons. Searches for semi-leptonic decays are often restricted to regions in $q^2$ away from the resonances to enhance sensitivity to NP as shown in Fig.~\ref{fig:brlongdist}. Forbidden decays refer to those which violate conservation of charged lepton flavor and lepton number. As resonance pollution to these decays is absent, no $q^2$ binning is needed for forbidden decay modes, and their studies represent a clear null test, complementary to those presented in Sec.~\ref{sec:nulltests}. Searches for decays of \Dz and \Dp mesons violating conservation of baryon number also exist\cite{Ablikim:2019lil,Rubin:2009aa}, but will not be discussed further in this review.

The most recent experimental results have been obtained by the \lhcb, \babar, \belle and \bes3 collaborations. The LHCb detector\cite{Alves:2008zz,Aaij:2014jba} is a single arm forward spectrometer designed to study decays of mesons containing a \cquark or \bquark quark, sited at the LHC (CERN, Switzerland). \lhcb has been designed to study proton--proton (\proton\proton) collisions in its main operation mode. \babar\cite{Aubert:2001tu} and \belle\cite{Abashian:2000cg} are cylindrical large-solid-angle detectors that operated at the PEP-II (SLAC National Accelerator Laboratory, USA) and KEKB (KEK, Japan) asymmetric-energy \ep\en colliders, known as \textit{b-factories}.
BES~III\cite{Ablikim:2009aa} is a general-purpose detector recording\break \ep\en collisions in the double-ring collider BEPCII (IHEP, China). Thanks to the large production cross-section at hadron colliders in the forward region\cite{Aaij:2013mga,Aaij:2015bpa}, \lhcb can profit from the world's largest recorded data set of charm hadron decays to date. Conversely, the other experiments benefit from detectors with excellent capabilities to reconstruct final states including neutral particles and electrons.

Older results published by \cleo II and the Fermilab E653 and E791 collaborations still hold the most stringent limits for some decay channels, see Refs.~\refcite{Freyberger:1996it,Kodama:1995ia,Aitala:1999db,Aitala:2000kk} for details. These measurements are not discussed in detail in this review, however, their results will be added to the summary tables for completeness. All experimental limits in this section are quoted at a 90\% confidence level (CL).

\subsection{Searches for purely leptonic decays}
Hadronic uncertainties on theoretical predictions are minimal in purely leptonic decays. However, their decay rates are subject to an additional helicity suppression, making them extremely rare. In SM, these decays are dominated by long-distance contributions from
$D^0\to \gamma^* \gamma^* \to \ell^+\ell^-$ (Refs.~\refcite{Burdman:2001tf,Petrov:2016kmb,Petrov:2017nwo}) with an estimated branching fraction of order $10^{-11}$ for muons, see Sec.~\ref{sec:modelindep}. Experimentally, the detection of final states consisting of two leptons is rather simple, generally allowing to set more stringent limits on the branching fractions compared to decays to final states involving hadrons. The possibility of NP searches in rare charm decays triggered the attention of experimental particle physicists already in 1988 and first searches for the rare decays \Dee, \Dmm and the lepton flavor violating (LFV) decay \Demu started by the \cleo and \argus collaborations\cite{Haas:1988bh,Albrecht:1988ge}. The world's most stringent limit nowadays on \Dee decays has been set by \belle in 2010, analyzing a dataset corresponding to an integrated luminosity of 660\invfb collected at a center-of-mass energy at or close to the \Y4S resonance\cite{Petri2010}. The best limits on the decays \Dmm and \Demu have been published by \lhcb in the years 2013\cite{Aaij:2013cza} and 2016\cite{Aaij:2015qmj}, respectively, using datasets corresponding to integrated luminosities of 0.9\invfb and 3.0\invfb.
The best upper limits on the branching fractions of purely leptonic rare charm decays are\cite{Petri2010,Aaij:2013cza,Aaij:2015qmj}
\begin{align} \label{eq:limitsDll}
\begin{split}
\BF(\Dee)		& < 7.9 \times 10^{-8}~,  \\
\BF(\Dmm)	&< 6.2 \times 10^{-9}~, \\
\BF(\Demu)	&< 1.3 \times 10^{-8}~.
\end{split}
\end{align}
%

\subsection{Searches for semi-leptonic decays}\label{sec:SLdecays}
Searches for semi-leptonic decays of neutral \Dz and charged \Dp, \Dsp charm mesons into two leptons and additional hadrons cover a large variety of final states, and we will briefly discuss the most recent publications. Singly Cabibbo-suppressed decays with two oppositely charged leptons $\ell^+\ell^-\,(\ell=e,\mu)$ in the final state of the form $D\to F\ell^+\ell^- $ are sensitive to FCNC processes. Here, $F$ can be one or several neutral ($\piz$, $\KS$, $\eta$, $\rhoz$, $\omegaz$, $\Kstarz$, $\phi$) and/or charged ($\pip$, $\Kp$, $\rhop$, $\Kstarp$) mesons.
Final states comprising two leptons of different flavor correspond to LFV modes, while modes with two leptons carrying the same electrical charge are lepton number violating (LNV). To date, no indications of non-resonant short-distance contributions to rare decay modes or hints for forbidden modes exist.

In autumn 2020, the \lhcb collaboration has published a search for 25 rare and forbidden decays of \Dp and \Ds mesons into two leptons and a charged kaon or pion, analyzing a data set of \proton\proton collisions corresponding to an integrated luminosity of 1.6\invfb\cite{Aaij:2020wyk}. The achieved limits in the range $(1.4\hbox{--}640) \times 10^{-8}$ improve the previous ones in most cases by at least one order of magnitude. Resonant contributions are minimized by vetoing the region $[525,1250]\mevcc$ in dilepton mass and extrapolating the signal yields to the vetoed regions assuming a uniform distribution of the particles across the phase space. The resonant decay modes proceeding via an intermediate $\phi$ meson $\D^+_{(s)} \to \pip [\ell^+\ell^-]_\phi$ are used as calibration and normalization.
The most stringent results are obtained analyzing final states involving muons.
Decay channels of \Dp mesons including negatively charged kaons such as
$\Dp\to\Km\ep\ep$, $\Dp\to\Km\mup\mup$, $\Dp\to\Km\mup\ep$ have not been investigated due to a large amount of background coming from misidentified $\Dp\to\Km\pip\pip$ decays. A listing of the observed limits can be found in Table~\ref{tab:ThreeBodyDecays}. The table summarizes\- the most stringent limits on rare and forbidden decay channels of \Dp and \Ds mesons that have been investigated to date.

\begin{table}[t]
\tbl{Most stringent upper limits (UL) at a 90~\% CL on the branching fractions of (left) rare and (right) forbidden decays of \Dp and \Ds mesons.}{
\resizebox{\textwidth}{!}{
\begin{tabular}{@{}l r @{/} l c | l r @{/} l c@{}} \toprule
 	Final state		&    \multicolumn{2}{c}{ \BF$[10^{-6}]$ (UL)} & Ref.	&	 	Final state		&     \multicolumn{2}{c}{\BF$[10^{-6}]$ (UL)} & Ref.	   \\		
	\hspace{0.5cm}$F$							&    	     	$\Dp$ & $\Dsp \to F$ 	&		& 	\hspace{0.5cm}$F$	&		$\Dp$ & $\Dsp \to F$	 		&		\\		\colrule	 
$\pip\, \epem$   	   	& 	1.1&5.5 	 	&BaBar\cite{Lees:2011hb}/LHCb\cite{Aaij:2020wyk} 				& 	$\pip \,\ep \mun$   	 	&	0.21&1.1	&LHCb\cite{Aaij:2020wyk} 	  \\
$\Kp\, \epem$	   	     & 		0.85&3.7	&LHCb\cite{Aaij:2020wyk}/BaBar\cite{Lees:2011hb} 	&	$\pip \,\mup \en$   	 	&  	0.22&0.94	&LHCb\cite{Aaij:2020wyk} 	  \\				
$\pip\, \mumu$   	  	& 		0.067&0.18&LHCb\cite{Aaij:2020wyk} 					&	$\pip \,\epm \mump$  &	 	34&610	&E791\cite{Aitala:1999db}  			 \\
$\Kp\, \mumu$    	    & 		 0.054&0.14&LHCb\cite{Aaij:2020wyk} 	    			&	$\Kp \,\ep \mun$   	    & 		0.075&0.79	&LHCb\cite{Aaij:2020wyk} 		 \\		
$\pip \,\piz \,\epem$  & 		 14&-- &BES III \cite{Ablikim:2018gro}	        	&$\Kp \,\mup \en$   	    & 	0.1&0.56	&LHCb\cite{Aaij:2020wyk} 	 \\			
$\Kp \,\piz\, \epem$  & 		 15&-- 	 &BES III \cite{Ablikim:2018gro}	&$\Kp\, \epm \mump$  	& 		68&630	&E791\cite{Aitala:1999db} 	 \\			
$\KS \,\pip\, \epem$  & 		 26&-- 	 &BES III \cite{Ablikim:2018gro} 	& 	 $\pim\, \ep \ep$   	  	& 	0.53&1.4	&LHCb\cite{Aaij:2020wyk} 	\\			
$\KS\, \Kp\, \epem$   & 		 11&--	 	 &BES III \cite{Ablikim:2018gro} 	& 	 $\Km \,\ep \ep$   	  	& 	0.9&0.77 &BaBar\cite{Lees:2011hb}/LHCb\cite{Aaij:2020wyk} \\			
$\rhop\, \mumu$    	 & 		 560&-- 	 	 &E653\cite{Kodama:1995ia} & $\pim\, \ep \mup$       	& 	0.13&0.63	&LHCb\cite{Aaij:2020wyk} 	 \\		  			
$\Kstarp\, \mumu$  & 		 	--&1400 	 &E653\cite{Kodama:1995ia} &$\Km \ep \mup$       	& 	1.9&0.26	&BaBar\cite{Lees:2011hb}/LHCb\cite{Aaij:2020wyk}  	 \\		
 \multicolumn{3}{c}{}{}{} 								& 								 & 	 $ \pim\, \mup \mup$ & 0.014&0.086		&LHCb\cite{Aaij:2020wyk} 			\\			
 \multicolumn{3}{c}{}{}{} 								& 	 							 & 	 $\Km \,\mup \mup$   & 10&0.026&BaBar\cite{Lees:2011hb} /LHCb\cite{Aaij:2020wyk} 	\\		
 \multicolumn{3}{c}{}{}{} 								& 								 &	$\KS \,\pim\, \ep \ep$ &  3.3& -- 	 & BES III\cite{Ablikim:2019gvd} 	 \\		
 \multicolumn{3}{c}{}{}{} 								& 								 &		$\Km\, \piz\, \ep \ep$ &  8.5& -- 	 & BES III\cite{Ablikim:2019gvd} 	 \\			
 \multicolumn{3}{c}{}{}{} 								& 							 & 	$\rhom \,\mup \mup$ & 		 560& -- 	 	 & E653\cite{Kodama:1995ia}	 \\		  			
 \multicolumn{3}{c}{}{}{} 								& 							 &		$\Kstarm \,\mup \mup$ &  850&1400 	 & E653\cite{Kodama:1995ia} 	 \\		
\botrule
\end{tabular} \label{tab:ThreeBodyDecays}}
}
\end{table}

The \bes3 collaboration has published a search for numerous decay channels of \Dp and \Dz mesons into final states comprising two electrons in 2018\cite{Ablikim:2018gro}. These are decays of charged \Dp mesons to two electrons accompanied by a pair of a neutral and a charged pseudoscalar meson (e.g.~$\Dp\to\piz\pip\ep\en$), decays of neutral \Dz mesons into two electrons plus a neutral meson (e.g.~$\Dz\to\piz\ep\en$) or two oppositely charged pseudoscalars (e.g.~$\Dz\to\pip\pim\ep\en$). The analysis uses \ep\en collision data corresponding to an integrated luminosity of 2.93\invfb. The data has been recorded at a center-of-mass energy of $\sqrt{s}=3.773\gev$, which is close to the \Dp\Dm or \Dz\Dzb mass threshold. Upper limits in the range $(3$--$41) \times 10^{-6}$ are determined using a double tagging approach\cite{PhysRevLett.56.2140,PhysRevLett.60.89} designed to measure absolute branching fractions.
Due to the limited luminosity and lower charm production cross section in \ep\en annihilations, the achieved limits are less stringent compared to those set by \lhcb on muonic modes\cite{Aaij:2013uoa,Aaij:2017iyr}.
The measured limits for all channels under study are summarized in Table~\ref{tab:ThreeBodyDecays} for decays of charged \Dp mesons, and limits on decays of neutral \Dz mesons are listed in Table~\ref{tab:FourBodyDecays}. In addition, a search for heavy Majorana neutrino LNV decays with two electrons has been published by \bes3 in 2019\cite{Ablikim:2019gvd}, where best limits on
$\Dp\to\KS\pim\ep\ep$ and $\Dp\to\Km\piz\ep\ep$ have been reported, which can also be found in Table~\ref{tab:ThreeBodyDecays}.

The most stringent upper limits on forbidden decays of neutral \Dz mesons have been reported by the \babar collaboration in two successive publications\cite{Lees:2019pej,Lees:2020qgv} during 2020. A data set of \ep\en annihilation corresponding to integrated luminosity 468\invfb recorded at or close to the $\Y4S$ resonance has been analyzed. The upper limits are set relative to decays of purely hadronic decays. Upper limits in the order of $(1.0\hbox{--}30.6) \times 10^{-7}$ and $(5.0\hbox{--}42.8 )\times 10^{-7}$ are found for decay modes involving a pair of oppositely charged and neutral hadrons, respectively. The achieved results can be found in Table~\ref{tab:FourBodyDecays}. With the exception of $\Dz\to\epm\mump$, all best limits on forbidden \Dz meson decays are set by \babar to date.

Note that more than half of the limits summarized in Tables~\ref{tab:ThreeBodyDecays} and \ref{tab:FourBodyDecays} originate from measurements published during 2018--2020, proving the great experimental progress that the field has made in recent times.

\begin{table}[tb]
\tbl{Most stringent upper limits (UL) at a 90~\% CL on the branching fractions of (left) rare and (right) forbidden decays of \Dz mesons.}
{\begin{tabular}{@{}lcc|lcc@{}} \toprule
     Final state				& 		          \BF$[10^{-6}]$ (UL) & Ref.	&	 	Final state	& 		                 \BF$[10^{-6}]$ (UL) & Ref.			  \\		
	\hspace{0.5cm}$F$		    &    	     	 $\Dz \to F$ 	        &		&		\hspace{0.5cm}$F$			&		  $\Dz \to F$	 		 &		\\		\colrule
$\ep \en$   	      & 	0.079  	&	 Belle	\cite{Petri__2010}	    &	$\epm \mump$   	        &  0.013        & LHCb\cite{Aaij:2015qmj}   \\
$\mup \mun$   	   	  & 	0.0062	&  LHCb\cite{Aaij:2013cza}	        & 	$\piz \epm \mump$ 	    &      0.80     & BaBar\cite{Lees:2020qgv}  \\
$\piz \,\epem$   	   	  & 	4	  	& 	BES III\cite{Ablikim:2018gro}    &  $\eta \,\epm \mump$        &      2.25     & BaBar\cite{Lees:2020qgv}   \\
$\eta \,\epem$   	   	  & 	3	    &	BES III\cite{Ablikim:2018gro}    &  $\rhoz\, \epm \mump$       &      0.50     & BaBar\cite{Lees:2020qgv}  \\
$\rhoz \,\epem$   	  & 	100	    &	CLEO II\cite{Freyberger:1996it} &  $\omega \,\epm \mump$      &      1.71     & BaBar\cite{Lees:2020qgv}  \\
$\omega \,\epem$   	  & 	6	    &	BES III\cite{Ablikim:2018gro}    &  $\phi \,\epm \mump$        &      0.51     & BaBar\cite{Lees:2020qgv}  \\
$\phi \,\epem$   	   	  & 	52	    &	CLEO II \cite{Freyberger:1996it}&  $\KS \,\epm \mump$         &      0.87     & BaBar\cite{Lees:2020qgv}  \\
$\Kzb \,\epem$   	   	  & 	110	    &	CLEO II\cite{Freyberger:1996it}	&  $\Kstarzb\, \epm \mump$    &      1.25     & BaBar\cite{Lees:2020qgv}  \\
$\KS \,\epem$   	   	  & 	12	    &	BES III\cite{Ablikim:2018gro}	&  $\pip \,\pim \,\epm \mump$   &      1.71     & BaBar\cite{Lees:2019pej}  \\
$\Kstarzb\, \epem$	  & 	47	  	&	E791\cite{Aitala:2000kk}      	&  $\Kp\, \Km\, \epm \mump$     &      1.00      & BaBar\cite{Lees:2019pej}  \\
$\piz \,\mumu$   	   	  & 	180	  	&	E653\cite{Kodama:1995ia}		&  $\Km \,\pip\, \epm \mump$    &      1.90      & BaBar\cite{Lees:2019pej} \\
$\eta\,\mumu$   	   	  & 	530	  	&	CLEO II\cite{Freyberger:1996it} &  $\pim \,\pim\, \ep \ep$  &      0.91     & BaBar\cite{Lees:2019pej} \\
$\rhoz \,\mumu$   	  & 	22	  	&	E791\cite{Aitala:2000kk}		&  $\Km\, \Km \, \ep \ep$   &      0.34     & BaBar\cite{Lees:2019pej} \\
$\omega\,\mumu$   	  & 	830	  	&	CLEO II\cite{Freyberger:1996it}	&  $\Km \,\pim\,  \ep \ep$  &     0.50      & BaBar\cite{Lees:2019pej} \\ 
$\phi \,\mumu$   	   	  & 	31	 	&	E791\cite{Aitala:2000kk}  		&  $\pim \,\pim\, \mup \mup$&     1.52      & BaBar\cite{Lees:2019pej}  \\
$\Kzb \,\mumu$   	   	  & 	260	 	& 	E653\cite{Kodama:1995ia}  		&  $\Km \,\Km\,  \mup \mup$ &     0.10      & BaBar\cite{Lees:2019pej}  \\  
$\Kstarzb \,\mumu$      & 	24	    &	E791\cite{Aitala:2000kk}  		&  $\Km \,\pim\,  \mup \mup$&     0.53      & BaBar\cite{Lees:2019pej}  \\
$\pip \,\pim \,\epem$     & 	7	  	&	BES III\cite{Ablikim:2018gro}    &  $\pim \,\pim\, \ep \mup$ &    3.06       & BaBar\cite{Lees:2019pej}	\\ 
$\Kp \,\Km\, \epem$   	  & 	11	&	BES III\cite{Ablikim:2018gro}	&  $\Km \,\Km\,  \ep \mup$  &     0.58      & BaBar\cite{Lees:2019pej}  \\
$\Km \,\pip\, \epem$   	  & 	41  	&	BES III\cite{Ablikim:2018gro}	&  $\Km \,\pim\,  \ep \mup$ &     2.10       & BaBar\cite{Lees:2019pej}	\\
$\pip \,\pim\, \mumu$     & 	0.55  	&	LHCb\cite{Aaij:2013uoa}  		&   										&					\\
$\Kp\, \Km\, \mumu$   	  & 	33  	&	E791\cite{Aitala:2000kk}		& 			    							 &				   \\
$\Km\, \pip\, \mumu$   	  & 	359	  	&	E791\cite{Aitala:2000kk}		& 				   							 &		    	    \\
$\piz \,\pip\, \pim\, \mumu$& 	810		&	E653\cite{Kodama:1995ia}  	    & 	  										 &			        \\
\botrule
\end{tabular}  \label{tab:FourBodyDecays}}
\end{table}

\subsection{Observation of resonance-dominated semi-leptonic decays}\label{sec:RasonantDecays}
Instead of vetoing resonance-dominated regions of the decay phase space, a complementary approach has recently been applied for $\Dz \to P_1 P_2 \ell^+ \ell^-$ decays, where $P_1 P_2$ stands for a pair of pseudoscalar mesons. The branching fraction measurements are done binned in regions of $q^2$. While limited sensitivity to the short-distance component is still given in bins which are away from the resonances, the regions around the resonances are fully dominated by long-distance contributions. However, signal decays in those bins can be used to perform in-depth studies of SM null tests as will be explained later in Sec.~\ref{sec:nulltests}.

The observation of the decay mode \Dkpmm has been reported by \lhcb in 2016\cite{Aaij:2015hva}. The measurement is limited to a region of dimuon mass of $[675\hbox{--}875]\mevcc$ around the $\rhoz/\omegaz$ mass, where a significant signal is observed for the first time. The branching fraction is measured relative to $\Dz \to \Km\pip\pim\pip$ decays to be\cite{Aaij:2015hva}
\begin{equation} 
\BF(\Dkpmm)=(\BFkpmm\pm\BFkpmmStat \pm \BFkpmmSyst)\BFkpmmUnit~,\\
\end{equation}
where the uncertainties are statistical and systematic, respectively. Since the decay is a Cabibbo-favored mode, no $|\Delta c|=|\Delta u|=1$ FCNC contributions are involved. However, the decay provides an important reference channel for further measurements of \Dz meson decays to four-body final states and to test QCD methods.

In 2017, the \lhcb collaboration has searched for the rare decays \Dppmm and \Dkkmm using a data set of \proton\proton collision corresponding to 2\invfb\cite{Aaij:2017iyr}. The measurement is done in bins of dimuon mass defined around the intermediate resonances and a high- and low-mass region, where the influence of the resonances is minimal. For \Dppmm decays these regions are: (low-mass) $<$\,$525\mevcc$, $(\eta)\,525\hbox{--}565\mevcc$, $(\rhoz/\omegaz)\,565\hbox{--}950\mevcc$, $(\phi)\,950\hbox{--}1100\mevcc$, and (high-mass) $>$\,$1100\mevcc$. Due to the reduced phase space, no $\phi$ and high-mass regions are accessible for \Dkkmm decays, where the $\rhoz/\omegaz$ region marks the kinematic endpoint at 565\mevcc. Significant signal has been observed in all regions of dimuon mass except for the $\eta$ and high-mass regions for \Dppmm. The branching fractions are measured using \Dkpmm decays as normalization employing the measured branching fraction discussed earlier in this section\cite{PhysRevD.98.030001}. The achieved limits and branching fractions are summarized in Table~\ref{tab:hhmumuBF} and confirm the expected dominance of resonant contributions in the decay processes.

\begin{table}[t]
\tbl{Branching fractions in different ranges of dimuon mass, where the uncertainties are statistical, systematic and due to the limited knowledge of the normalisation branching fraction. Table from Ref.~\protect\refcite{Aaij:2017iyr}.}
{\begin{tabular}{@{}lr@{--}lc@{}} \toprule
\multicolumn{4}{c}{\Dppmm}\\
\hline
\mmumu region & \multicolumn{2}{c}{[$\mevcc$]} & $\BF$ [$10^{-8}$]\\
Low mass        & \multicolumn{2}{c}{$<525$} & $\enspace7.8\pm 1.9 \pm 0.5 \pm 0.8$\\
$\eta$          & 525 & 565  & $<2.4$\\
$\rhoz/\omegaz$ & 565 & 950  & $40.6\pm 3.3 \pm 2.1 \pm 4.1$\\
$\phi$          & 950 & 1100 & $45.4\pm 2.9 \pm 2.5 \pm 4.5$\\
High mass       & \multicolumn{2}{c}{$>1100$} & $<2.8$\\
\hline
\multicolumn{4}{c}{\Dkkmm}\\
\hline
\mmumu region & \multicolumn{2}{c}{[$\mevcc$]} & $\BF$ [$10^{-8}$]\\
Low mass        &\multicolumn{2}{c}{$<525$} & $\enspace2.6 \pm  1.2 \pm 0.2 \pm 0.3$\\
$\eta$          &525 & 565  & $<0.7$\\
$\rhoz/\omegaz$ &\multicolumn{2}{c}{$>565$} & $12.0 \pm 2.3  \pm 0.7  \pm 1.2$ \\
\botrule
\end{tabular} \label{tab:hhmumuBF}}
\end{table}

Integrating over the full dimuon-mass region, the total branching fractions of \Dppmm and \Dkkmm decays has been found to be consistent with SM expectations\cite{Cappiello:2012vg} and measured to be\cite{Aaij:2017iyr}
\begin{align}\begin{split}
\BF(\Dppmm)&=(\BFppmm\pm\BFppmmStat\pm\BFppmmSyst\pm\BFppmmNorm)\BFppmmUnit~,\\
\BF(\Dkkmm)&=(\BFkkmm\pm\BFkkmmStat\pm\BFkkmmSyst\pm\BFkkmmNorm)\BFkkmmUnit~,
\end{split}
\end{align}
where the uncertainties are statistical, systematic and due to the limited knowledge of the normalization branching fraction, respectively. Even though dominated by resonant contributions, these represent the rarest observed decays of charmed mesons to date.

As of today, a single decay mode with two electrons in the final state has been observed. The \babar collaboration succeeded in making the first observation of the decay \Dkpee in a region of invariant mass of the electron-positron pair of [675--875]\mevcc around the $\rhoz/\omegaz$ resonances\cite{Lees:2018vns}, coinciding with the dimuon-mass region defined by \lhcb when measuring the muonic mode\cite{Aaij:2017iyr}. The analysis uses a data set of \ep\en annihilation data corresponding to 468\invfb. Fully hadronic decays of $\Dz\to\Km\pip\pim\pip$ decays are used as normalization. The branching fraction has been determined to be\cite{Lees:2018vns}
\begin{equation} 
\BF(\Dkpee)=(\BFkpee\pm\BFkpeeStat \pm \BFkpeeSyst \pm \BFkpeeNorm)\BFkpeeUnit~,
\end{equation}
where the uncertainties are statistical, systematic and due to the limited knowledge of the normalization branching fraction, respectively.
Assuming a SM rate, the available statistics accumulated by \babar, \belle and \bes3 is expected to be insufficient to make further observations of rare four-body dielectron modes. \lhcb has not yet started to measure branching fractions of resonance-dominated decays involving two electrons in the final state, however, contributions are expected in the near future\cite{Bediaga:2018lhg}.

\subsection{Studies of semi-leptonic baryonic decays}
The most recent searches for rare baryonic decays are restricted to decays of $\Lambda_c$ baryons decaying into a pair of leptons accompanied by a (anti)proton. While the world's best limits on the dielectron decay $\Lambda_c \to p\ep \en$, as well as LFV and LNV decays have been set in the order $(2.7-19.0)\times10^{-6}$ by \babar already in 2011\cite{Lees:2011hb}, \lhcb has investigated decays into two muons and a proton in the year 2018 using \proton\proton collision data corresponding to 3\invfb\cite{Aaij:2017nsd}. The new limit
\begin{equation} 
\BF(\Lambda_c \to \proton\, \mumu)<7.7 \times10^{-8}~,
\end{equation}
improves the previous result on the branching fraction in the non-resonant region of dimuon mass by more than two orders of magnitude. Ranges $\pm 40\mevcc$ around the known $\omega$ and $\phi$ meson masses have been vetoed. The result reveals the capability of \lhcb to make significant contributions in investigations of baryonic decays.
\begin{table}[t]
\tbl{Most stringent upper limits (UL) at a 90~\% CL on the branching fractions of (left) rare and (right) forbidden decays of $\Lambda_c$ baryons.}
{\begin{tabular}{@{}lcc|lcc@{}} \toprule
 	Final state	$F$		& 		          \BF$[10^{-6}]$ (UL) & Ref.	&	 	Final state	$F$	& 		                 \BF$[10^{-6}]$ (UL) & Ref.			  \\		
								&    	     	 $\Lambda_c \to F$ 	&											& 									&		  $\Lambda_c \to F$	 		 &		\\				\colrule
$ p \,\ep \en$   	   		 				& 5.5	  	&	 BaBar \cite{Lees:2011hb} 					& 	$  p \,\ep \mun$   	 & 	9.9		&  BaBar\cite{Lees:2011hb} 	   \\
$ p \,\mup \mun$   	   		 		& 	0.077	&  LHCb\cite{Aaij:2017nsd}						& 	$	 p\, \mup \en$   	 & 	19.0	  	&	BaBar \cite{Lees:2011hb}	    \\
                             		& 		  	& 	                            				&  $	\bar{p}  \,\ep \ep$ 		& 	2.7 	& 	BaBar\cite{Lees:2011hb}   \\
															   		& 	  		 	&										&  $\bar{p} \, \mup \mup$ & 	9.4& BaBar\cite{Lees:2011hb}    \\ 
															   		& 	  		 	&										&  $\bar{p}  \,\ep \mup$ & 	16.0& BaBar\cite{Lees:2011hb}    \\ 
	        												   		& 	  		 	&										&  $ \Sigma^{-} \, \mup \mup$  & 	700	  	& 	E653\cite{Kodama:1995ia}  \\ \botrule
\end{tabular}  \label{tab:baryonicDecays}}
\end{table}
In addition, a significant signal of $13.2\pm4.3$ candidates has been observed at the $\Lambda_c$ mass in the region of dimuon mass around the $\omega$ mass with a significance of $5\sigma$, which could open the possibility to study angular distributions in the resonance-dominated regions in the near future. The branching fraction of the decay $\Lambda_c \to \proton[\mup\mun]_\omega$ in the region [759,\,805]\mevcc of dimuon mass has been measured to be\cite{Aaij:2017nsd}
\begin{equation}
\BF(\Lambda_c \to \proton\, [\mup\mun]_\omega) =(9.4 \pm 3.2 \pm 1.0 \pm 2.0 ) \times 10^{-4},\\
\end{equation} 
where the uncertainties are statistical, systematic and due to the limited knowledge of the normalization branching fraction, respectively. The resonant decay mode proceeding through an intermediate $\phi$ meson $\Lambda_c \to \proton [\mup\mun]_\phi$ has been used as normalization, using the known branching fractions of the decays $\Lambda_c \to \proton\phi$ and $\phi \to \mup\mun$.

A single measurement investigating decays of $\Sigma_c$ baryons exists by the E653 collaboration\cite{Kodama:1995ia}, dating back to the year 1995. We give a summary of the most stringent limits on rare charmed $\Lambda_c$ baryon decay modes that have been studied to date in Table~\ref{tab:baryonicDecays}.

\section{Model-independent constraints on Wilson coefficients}\label{sec:modelindep}

Using the most recent available experimental results presented in the previous section, we report the most stringent constraints on Wilson coefficients in the charm system, separately assuming lepton-flavor conservation and allowing for LFV.

\subsection{Lepton flavor conserving bounds}\label{sec5.1}
Currently, the best constraints on dimuon and dielectron Wilson coefficients are obtained from upper limits on
$\mathcal{B}(D^+\to \pi^+ \mu^+\mu^-)$ and $\mathcal{B}(D^+\to \pi^+ e^+ e^-)$.
Using the experimental upper limits on $\mathcal{B}(D^+\to \pi^+ \mu^+\mu^-)<6.7\cdot 10^{-8}$ at 90\%~CL presented in Sec.~\ref{sec:expSearches}, and neglecting SM contributions, we obtain the following bounds for muons:
\begin{align}\label{eq:NP_constraints_muons}\begin{split}
1.3&\,\left|\mathcal{C}_7\right|^2 + 1.3\,\left|\mathcal{C}_9^{(\mu)}\right|^2 + 1.3\,\left|\mathcal{C}_{10}^{(\mu)}\right|^2 + 2.6\,\left|\mathcal{C}_S^{(\mu)}\right|^2 \\ 
+\, 2.7&\,\left|\mathcal{C}_P^{(\mu)}\right|^2 + 0.4\left|\mathcal{C}_T^{(\mu)}\right|^2 + 0.4\,\left|\mathcal{C}_{T5}^{(\mu)}\right|^2  + 0.3\,\mathrm{Re}\left[\mathcal{C}_9^{(\mu)} \,{\mathcal{C}_T^{(\mu)}}^*\right]\\
+\, 1.1&\,\mathrm{Re}\left[\mathcal{C}_{10}^{(\mu)}\, {\mathcal{C}_P^{(\mu)}}^*\right] + 2.6\,\mathrm{Re}\left[\mathcal{C}_7 \,{\mathcal{C}_9^{(\mu)}}^*\right] + 0.6\,\mathrm{Re}\left[\mathcal{C}_7\, {\mathcal{C}_T^{(\mu)}}^*\right] \lesssim1 \,,
\end{split}
\end{align}
Note that coefficients in Eq.~\eqref{eq:NP_constraints_muons} are slightly improved with respect to Ref.~\refcite{Bause:2019vpr}. Right-handed currents can be included replacing $\mathcal{C} \to \mathcal{C} + \mathcal{C}^{\prime}$.

The decay $D^0\to \mu^+\mu^-$ provides additional information on $\mathcal{C}^{(\mu)(\prime)}_{S,P,10}$ via the experimental upper bound presented in Eq.~\eqref{eq:limitsDll}. The short-distance contributions can be written as
\begin{equation}
\begin{split}
\mathcal{B}(D^0\to\mu^+\mu^-) &= {G_F^2\, \alpha_e^2 \,m_D^5\, f_D^2 \over 64\,\pi^3\, m_c^2\, \Gamma_D} \sqrt{1 - {4\,m_\mu^2 \over m_D^2}} \left[ \left( 1 - {4\,m_\mu^2 \over m_D^2} \right) \left|\mathcal{C}^{(\mu)}_S - {\mathcal{C}^{(\mu)}_S}^\prime\right|^2 \right.\\
& \quad \left.+\left|\mathcal{C}^{(\mu)}_P - {\mathcal{C}^{(\mu)}_P}^\prime + {2m_\mu m_c \over m_D^2} \left(\mathcal{C}^{(\mu)}_{10} - {\mathcal{C}^{(\mu)}_{10}}^\prime\right) \right|^2 \right] \,,
\end{split}
\label{eq:Br_D-mumu}
\end{equation}
where $f_D$ is the $D^0$ meson decay constant, see Sec.~\ref{sec:FFs}. Albeit the active helicity suppression in Eq.~\eqref{eq:Br_D-mumu}, it gives the best constraints on NP contributions to ${\mathcal{C}^{(\mu)}_{10}}^{(\prime)}$,\cite{Bause:2019vpr}
\begin{align}\label{eq:Dmumu}
    \left|\mathcal{C}^{(\mu)}_S - {\mathcal{C}^{(\mu)}_S}^\prime\right|^2 +\left|\mathcal{C}^{(\mu)}_P - {\mathcal{C}^{(\mu)}_P}^\prime + 0.1\,\left(\mathcal{C}^{(\mu)}_{10} - {\mathcal{C}^{(\mu)}_{10}}^\prime\right) \right|^2\lesssim 0.007\,,
\end{align}
where the SM contributions have again been neglected since the relevant Wilson coefficients are zero in the SM, as discussed in Sec.~\ref{sec:shortdist}. The most sizeable contribution in the SM is estimated to originate from long-distance contributions in $D^0\to \gamma^* \gamma^* \to \mu^+\mu^-$\cite{Burdman:2001tf,Petrov:2016kmb,Petrov:2017nwo}. Using the current limit $\mathcal{B}(D^0\to \gamma \gamma)<8.5\cdot 10^{-7}$ at 90\%~CL\cite{Tanabashi:2018oca}, we obtain\cite{Burdman:2001tf}
\begin{align}
\label{eq:smcontrellell}
    \mathcal{B}(D^0\to \mu^+\mu^-)_{\rm LD}\approx 8\,\alpha^2\cdot\left(\frac{m_\mu^2}{m_D^2}\right)\cdot\text{log}^2\left(\frac{m_\mu^2}{m_D^2}\right) \cdot \mathcal{B}(D^0\to \gamma \gamma) \sim 10^{-11}\,,
\end{align}
well below the current experimental limit (see Eq.~\eqref{eq:limitsDll}).
Equation~\eqref{eq:smcontrellell} can be considered as an upper limit and any measurement significantly exceeding this bound would signal NP.

Constraints on dielectron modes are weaker than dimuon ones, using the upper limits on
$\mathcal{B}(D^+\to \pi^+ e^+ e^-)<1.1\cdot 10^{-6}$ and
$\mathcal{B}(D^0\to e^+ e^-)<7.9\cdot 10^{-8}$ at 90\%~CL presented in Sec.~\ref{sec:expSearches}, one obtains\cite{deBoer:2015boa}
\begin{align}\label{eq:electronbounds}
\begin{split}
    &\left|\mathcal{C}^{(e)}_{S,P} - {\mathcal{C}^{(e)}_{S,P}}^\prime\right|\lesssim 0.3~,\\
    &\left|\mathcal{C}^{(e)}_{9,10} - {\mathcal{C}^{(e)}_{9,10}}^\prime\right|\lesssim 4~,\quad \left|\mathcal{C}^{(e)}_{T,T5}\right|\lesssim 5~,\quad \left|\mathcal{C}_7\left({\mathcal{C}^{(e)}_{9}}-{\mathcal{C}^{(e)}_{9}}^\prime\right)\right|\lesssim 2~,
\end{split}
\end{align}
which are approximately a factor three weaker than the muonic ones, see Eqs.~\eqref{eq:NP_constraints_muons} and \eqref{eq:Dmumu}. Similar bounds on muon and electron Wilson coefficients are obtained with the study of the transverse momentum ($p_T$) spectrum of dilepton pairs produced in $\proton\proton$ collisions, recorded by the ATLAS\cite{Collaboration_2008} and CMS\cite{Chatrchyan:1129810} detectors. In particular, the high-$p_T$ range of the spectrum can be used to determine bounds on Wilson coefficients. Due to the limited phase space of charm decays, bounds on $\tau$ Wilson coefficients are only accessible using high-$p_T$ data\cite{Fuentes-Martin:2020lea} and the current bounds read,
\begin{align}\label{eq:taubounds}
\begin{split}
    &\left|{\mathcal{C}^{(\tau)(\prime)}_{9,10}}\right|\lesssim 6~,\,\quad\left|{\mathcal{C}^{(\tau)}_{T,T5}}\right|\lesssim 5~,\,\quad\left|{\mathcal{C}^{(\tau)(\prime)}_{S,P}}\right|\lesssim 14~.
\end{split}
\end{align}

In summary, Eqs.~\eqref{eq:NP_constraints_muons}, \eqref{eq:Dmumu}, \eqref{eq:electronbounds} and \eqref{eq:taubounds} represent the strongest constraints on Wilson coefficients for lepton flavor conserving transitions to date.

\subsection{Lepton flavor violating bounds}\label{sec5.2}
The phenomenology of these decays can be adopted from the lepton flavor conserving case, by introducing lepton flavor indices for operators and different Wilson coefficients. For clarity, we additionally indicate Wilson coefficients corresponding to LFV operators with ${K}$. We update the bounds on these LFV Wilson coefficients\cite{Bause:2019vpr}
\begin{equation}
\begin{split}
& 0.4\,|K_9|^2 + 0.4\,|K_{10}|^2 + 0.9\, |K_S|^2 + 0.9\,|K_P|^2  + 0.1\, |K_T|^2 + 0.1\, \big|K_{T5}\big|^2  + \\
& 0.2\,\mathrm{Re}\big[ K_{10} K_P^* \pm K_9 K_S^* \big]  + 0.1\,\mathrm{Re}\big[ K_9 K_T^* \pm K_{10} K_{T5}^* \big] \lesssim 1 \, ,
\end{split}
\label{eq:constraint_D-pimue}
\end{equation}
where $K_i=K_i^{e\mu} + K_i^{\prime e\mu}$ for $D^+\to \pi^+ e^+\mu^-$ and $K_i=K_i^{\mu e} + K_i^{\prime\mu e}$ for $D^+\to \pi^+ \mu^+e^-$.
$K_{S,P}^{(\prime)}$ can be better constrained from data on leptonic decays, using the measurement of $\mathcal{B}(D^0\to e^\pm\mu^\mp)$ presented in Eq.~\eqref{eq:limitsDll}. One obtains
\begin{equation}
\big|K_S - K_S^\prime \pm 0.04\,\big( K_9- K_9^\prime \big)\big|^2 + \big|K_P - K_P^\prime + 0.04\,\big( K_{10} - K_{10}^\prime \big)\big|^2 \lesssim 0.01 \, . \\
\label{eq:constraint_D-mue}
\end{equation}

As stated in Ref.~\refcite{Bause:2019vpr}, $K^{e\tau}$ couplings can be constrained by measurements of $\mathcal{B}(D^0\to e^{\pm} \tau^{\mp})$. However, no experimental measurement exists to date and only high-$p_T$ constraints are available for $K^{\mu\tau}$ and $K^{e\tau}$, see for instance Ref.~\refcite{Angelescu:2020uug}.

\section{Charming opportunities to probe the Standard Model with null tests} \label{sec:nulltests}

Due to its unique properties, the charm system allows the definition of clean observables which are null in the SM. In the following section, we outline ways to search for NP by testing for lepton universality, investigating angular distributions, CP asymmetries and radiative decays.

\subsection{Testing lepton universality}\label{sec:LU} 

In the last years, lepton universality (LU) ratios have gained interest due to anomalies found in the beauty sector\cite{Bifani:2018zmi}. The corresponding ratios in the charm sector are defined as\cite{Fajfer:2015mia,deBoer:2015boa,Bause:2019vpr,deBoer:2018buv}
\begin{align}
 R_{F}^{h_c} = \frac{\int_{q^2_{\text{min}}}^{q^2_{\text{max}}} \frac{\text{d}\mathcal{B}(h_c\to F\mu^+\mu^-)}{\text{d}q^2} \text{d}q^2 }
{\int_{q^2_{\text{min}}}^{q^2_{\text{max}}} \frac{\text{d}\mathcal{B}(h_c\to F e^+ e^-)}{\text{d}q^2} \text{d}q^2 }.
\end{align}
In the SM, the electroweak gauge bosons couple to leptons of different generations with equal strength, leading to a well-controlled SM prediction of\cite{Bause:2019vpr,deBoer:2018buv}
\begin{align}
\label{eq:SMLUratio}
(R_{F}^{h_c})_{\text{SM}}= 1\pm\mathcal{O}(\%).
\end{align}
Phase--space and electromagnetic corrections to $R_{F}^{h_c}$ are highly suppressed and cannot exceed the percent level\cite{Huber:2005ig,Bobeth:2008ij,Bordone:2016gaq}.
NP extensions do not necessarily respect LU, so $R_{F}^{h_c}$ can significantly deviate from unity. These ratios greatly profit from the cancellation of the dominant theoretical and experimental uncertainties if the same kinematic cuts ($q^2_{\text{max}}$, $q^2_{\text{min}}$) are applied to the dielectron and dimuon modes\cite{Hiller:2003js}. As an example, Table~\ref{tab:R_ratios} indicates experimentally possible NP values of $R_{\pi^+}^{D^+}$ and $R_{K^+}^{D_s^+}$ for the full, low and high $q^2$-integrated intervals\cite{Bause:2019vpr}.
Notice that the largest BSM signals are expected in the low and high $q^2$ regions, where the SM contributions are smaller. However, due to the limited knowledge of the resonance contributions in the low $q^2$ region, an interpretation in terms of specific NP models is harder than in the high $q^2$ region, where resonance influences are minimal. BSM predictions of LU ratios in $D\to P_1P_2\ell^+\ell^-$ modes have been worked out in Ref.~\refcite{deBoer:2018buv}. Effects up to 15\% are found on $R^{\Dz}_{\pip\pim}$ in the $q^2$-integrated range for\break \Dppmm decays.
\begin{table}[t]
 \tbl{$R_{\pi^+}^{D^+}$ in the SM and in NP scenarios for different $q^2$ regions. Ranges correspond to uncertainties from form factors and resonance parameters.\cite{Bause:2019vpr}} {
 \resizebox{\textwidth}{!}{  
 \begin{tabular}{l c c c c c c c} \toprule
&SM & $\vert \mathcal{C}^{(\mu)}_9 \vert=0.5$& $\vert \mathcal{C}^{(\mu)}_{10}\vert=0.5$ & $\vert \mathcal{C}^{(\mu)}_{9}\vert=\pm\vert \mathcal{C}^{(\mu)}_{10}\vert=0.5$  & $\vert \mathcal{C}^{(\mu)}_{S\,(P)}\vert=0.1$ & $\vert \mathcal{C}^{(\mu)}_T \vert=0.5$&$\vert \mathcal{C}^{(\mu)}_{T5}\vert=0.5$\\
 \hline
  full $q^2$ & $1.00 \pm \mathcal O(\%)$ &  SM-like	& SM-like	& SM-like& SM-like& SM-like& SM-like\\
  low $q^2$ & $0.95 \pm \mathcal O(\%)$ & $\mathcal{O}(100)$		& $\mathcal{O}(100)$		& $\mathcal{O}(100)$&$0.9\ldots1.4$& $\mathcal{O}(10)$&$1.0\ldots5.9$\\
  high $q^2$ & $1.00 \pm \mathcal O(\%)$ & $0.2\ldots11$		& $3\ldots7$		& $2\ldots17$& $1\ldots2$&$1\ldots5$&$2\ldots4$\\ \botrule
 \end{tabular} \label{tab:R_ratios}} 
  }
\end{table}
Due to the lack of branching fraction measurements of dielectron decay modes, LU tests in $c\to u \ell^+ \ell^-$ processes are largely unexplored and only limits exist. Na\"ive LU ratios in Ref. \refcite{deBoer:2018buv} find bounds on $R^{\Dz}_{\pip\pim}\gtrsim 0.1$ and $R^{\Dz}_{\Kp\Km}\gtrsim 0.01$ using the available upper limits on the dielectron modes\cite{Ablikim:2018gro} and branching fraction measurements of the dimuon modes\cite{Aaij:2017iyr} discussed in Secs.~\ref{sec:SLdecays} and \ref{sec:RasonantDecays}, respectively.
As resonant-dominated dielectron modes are expected within the reach of the \lhcb future sensitivity\cite{Bediaga:2018lhg}, we expect first model-independent LU tests in rare charm decays in the near future.

\subsection{Angular observables}\label{sec:angularobs} 

In the SM, the absence of axial vector lepton currents in rare charm decays (corresponding to $\mathcal{C}_{10}^{\text{SM}}=0$, cf.~Eq.~\eqref{eq:WCsSMzero}) leads to very specific angular distributions of the final state particles, and allows for clean null tests irrespective of form factors and their uncertainties.

The theory predictions of form factors have improved in recent years and a variety of $|\Delta c|=|\Delta u|=1$ decay modes can be studied, see Table~\ref{tab:FFs}. For angular distributions the number of observables increases with number of particles in the final state, as well as with the spin of particles included. In that manner, studying angular observables of different decay modes helps to pin down combinations of NP Wilson coefficients in a complementary way. In the following, the angular distribution for key channels are studied. Further model-dependent studies are available in e.g.~Refs.~\refcite{Bharucha:2020eup,Paul:2011ar,Fajfer:2006yc,Bigi:2011em,Paul:2012ab} and \refcite{Sahoo:2017lzi}.

\subsubsection{$D\to P \,\ell^+ \ell^-$} \label{sec:angularDPll}

The differential distribution of $D\to P \ell^+ \ell^-$ transitions reads\cite{Bobeth:2007dw}
\begin{equation}
{\text{d}^2\Gamma \over \text{d}q^2\,\text{d}\cos\theta} = a(q^2) + b(q^2) \cos\theta + c(q^2) \cos^2\theta \,,
\end{equation}
where $\theta$ denotes the angle between the $\ell^-$--momentum and the $P$-momentum in the dilepton rest frame. Its particular angular distribution provides two clean null tests of the SM: the lepton forward--backward asymmetry $A_{\rm FB}$,
\begin{align}
\begin{split}
    A_{\rm FB}(q^2) &= {1 \over \Gamma} \left[ \int_0^1 - \int_{-1}^0 \right] {\text{d}^2\Gamma \over \text{d}q^2 \text{d}\cos\theta} \text{d}\cos\theta= {b(q^2) \over \Gamma}\,,
\end{split}
\label{eq:AFB}
\end{align}
and the ``flat'' term $F_H$,
\begin{equation}
\begin{split}
F_H(q^2) &= {2\over \Gamma}\left[a(q^2) + c(q^2)\right] \,,
\end{split}
\label{eq:FH}
\end{equation}
where
\begin{align} \label{eq:G}
\Gamma=\Gamma(q^2_{\rm min},q^2_{\rm max})= \int_{q^2_{\rm min}}^{q^2_{\rm max}} {\text{d}\Gamma \over \text{d}q^2}\text{d}q^2 = 2  \int_{q^2_{\rm min}}^{q^2_{\rm max}} \left(a(q^2) + \frac{ c(q^2)}{3}\right) \,  \text{d}q^2 \, . 
\end{align} 
with integration limits depending on the $q^2$-bin. These observables are sensitive to operators with Lorentz structures not present in the SM. Neglecting lepton masses, it follows that
\begin{align}\label{eq:AFBFH}
    \begin{split}
        A_{\rm FB}&\propto \text{Re}\left[ \mathcal{C}_S\, \mathcal{C}_T^* + \left( \mathcal{C}_P + \mathcal{C}_P^R \right) \mathcal{C}_{T5}^* \right]~,\\
        F_H&\propto|\mathcal{C}_S|^2  + |\mathcal{C}_P + \mathcal{C}_P^R|^2\;\&\;\,|\mathcal{C}_T|^2  + |\mathcal{C}_{T5}|^2~.
    \end{split}
\end{align}
Taking the bounds on Wilson coefficients from Eqs.~\eqref{eq:NP_constraints_muons}, \eqref{eq:Dmumu} and \eqref{eq:electronbounds}, one obtains\cite{deBoer:2015boa}
\begin{align}
\begin{split}
    |A_{\rm FB}(D^+\to \pi^+ \mu^+\mu^-)|&\lesssim0.6~,\quad F_H(D^+\to \pi^+ \mu^+\mu^-)\lesssim1.5~,\\
|A_{\rm FB}(D^+\to \pi^+ e^+ e^-)|&\lesssim0.8~,\quad\,\, F_H(D^+\to \pi^+ e^+ e^-)\lesssim1.6~.
\end{split}
\end{align}
$A_{\rm FB}$ is zero in the SM up to higher corrections, which are suppressed by powers of $m_D/M_W$ and come from higher-dimensional operators or by $\alpha_e/(4\pi)$ ($D\to\pi\gamma\gamma\to\pi\ell^+\ell^-$)\cite{Bobeth:2007dw,deBoer:thesis}.
$F_H$ is also highly suppressed in the SM. For instance, $F_H(D^+ \to \pi^+ \mu^+ \mu^-)\sim\mathcal{O}(10^{-3})$ at low $q^2$, whereas $F_H(D_s^+ \to K^+ \mu^+ \mu^-)$ is $\mathcal{O}(10^{-2})$, and both are even smaller at high $q^2$\cite{Bause:2019vpr}. In addition, $F_H(D_{(s)}^+ \to \pi^+(K^+) e^+ e^-)$ is even further suppressed in the SM. Then, any nonzero measurement of these observables hints to NP generated by (pseudo)-scalar and (pseudo)-tensor operators, see Eq.~\eqref{eq:AFBFH}.

\subsubsection{ $\Lambda_c\to p \,\ell^+ \ell^-$}

The observables $A_{\rm FB}$ and $F_H$ introduced for $D\to P \ell^+ \ell^-$ in Sec.~\ref{sec:angularDPll} can also be defined for baryonic decays $\Lambda_c\to p \ell^+ \ell^-$. Ref.~\refcite{Meinel:2017ggx} performs an analysis of $A_{\rm FB}$ and $F_H$ in $\Lambda_c\to p \mu^+ \mu^-$, considering only contributions from $\mathcal{C}_{7,9,10}^{(\mu)(\prime)}$. In contrast to $F_H(D\to P \ell^+ \ell^-)$, the SM prediction of $F_H(\Lambda_c\to p \mu^+ \mu^-)$ is polluted by resonance contributions (mainly from $\mathcal{C}_9^R$), leading to values of $F_H(\Lambda_c\to p \mu^+ \mu^-)_{\rm SM}\sim 0.6$, such that it cannot be considered as clean null test.

On the other hand, $A_{\rm FB}(\Lambda_c\to p\mu^+\mu^-)_{\rm SM}$ vanishes because it is proportional to the product $\mathcal{C}_{9}^{(\mu)}\mathcal{C}_{10}^{(\mu)}$. Consequently, a small NP contribution to $\mathcal{C}_{10}^{(\mu)}$
could be enhanced through the interference with the resonant contributions from $\mathcal{C}_{9}^{R}$, producing a nonzero value of $A_{\rm FB}(\Lambda_c\to p\mu^+\mu^-)$, causing a clear sign of NP.

\subsubsection{$\Dz\to P_1 P_2 \,\ell^+ \ell^-$}

Due to more particles in the final state compared with the previous examples, the description of $\Dz\to P_1 P_2 \ell^+ \ell^-$ requires additional kinematic parameters that consequently lead to further angular observables. The angular distribution of $\Dz\to P_1 P_2 \ell^+ \ell^-$ can be written as\cite{Bobeth:2008ij,Bobeth:2012vn,deBoer:2018buv}
\begin{eqnarray}   \label{eq:full}
\frac{\text{d}^5\Gamma}{\text{d}q^2\,\text{d}p^2\,\text{d}\cos\theta_{P_1}\,\text{d}\cos\theta_\ell\, \text{d}\phi} &=&\frac{1}{ 2\,\pi} \left[ \sum_{i=1}^9 c_i(\theta_\ell,\phi)\, I_i (q^2,p^2,\cos \theta_{P_1}) \right]\,,
\end{eqnarray}
where $q^2$ and $p^2$ denote the invariant mass-squared of the dileptons and ($P_1 P_2$)-subsystem, respectively, and
\begin{align} \label{eq:ang_basis}
c_1 & =1\,, \quad c_2=\cos 2\,\theta_\ell\,, \quad c_3=\sin^2\theta_\ell\cos 2\phi\,, \quad c_4=\sin 2\,\theta_\ell \cos \phi\,, \quad c_5=\sin\theta_\ell\cos\phi\,, \nonumber \\ c_6& =\cos\theta_\ell\,, \quad c_7=\sin\theta_\ell\sin\phi\,, \quad c_8=\sin 2\,\theta_\ell\sin\phi\,, \quad c_9=\sin^2\,\theta_\ell\sin2\phi \, .
\end{align}
Here, $\theta_\ell$ is the angle between the $\ell^+$-momentum and the $\Dz$-momentum in the dilepton center-of-mass system (cms), while $\theta_{P} $ denotes the angle between the $P^+$-momentum and the negative direction of flight of the $\Dz$-meson in the ($P_1P_2$)-cms, and $\phi$ is the angle between the normal of the ($P_1P_2$)-plane and the ($\ell^+\ell^-$)-plane in the $D^0$ rest frame.
The angular coefficients $I_i \equiv I_i(q^2,p^2,\cos \theta_{P_1})$ in terms of transversity amplitudes can be found in Ref.~\refcite{deBoer:2018buv}. In particular, since $\mathcal{C}_{10}^{\text{SM}(\prime)}=0$, $I_{5,6,7}$ are zero in the SM. Hence, they constitute a formidable set of null tests observables\cite{deBoer:2018buv,Cappiello:2012vg}. From the CP-odd angular coefficients $I_{5,6,8,9}$ , asymmetries can be defined as
\begin{align}\label{eq:Ai}
A_i=\frac{I_i-\bar{I_i}}{\Gamma_{\text{ave}}}
\end{align}
with the CP-averaged decay rate $\Gamma_{\text{ave}}=\frac{\Gamma+\bar{\Gamma}}{2}$ and the angular coefficients ${I_i}$($\bar{I_i}$) for \Dz(\Dzb) mesons. In the SM, the observables $A_{5,6,8,9}$ are negligible given the experimental sensitivities today and constitute (approximate) SM null tests\cite{deBoer:2018buv,Cappiello:2012vg}.

\begin{figure}[t]
\begin{center}
\includegraphics[width=0.43\textwidth]{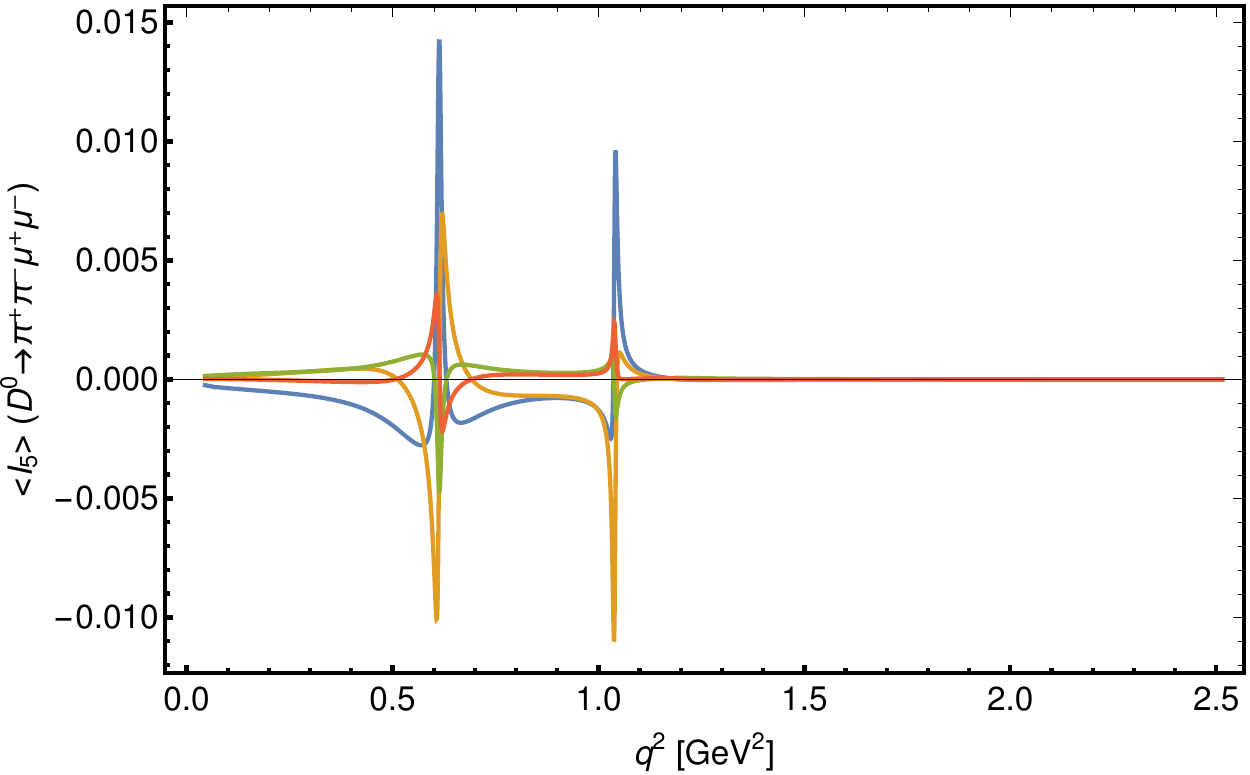}
\includegraphics[width=0.54\textwidth]{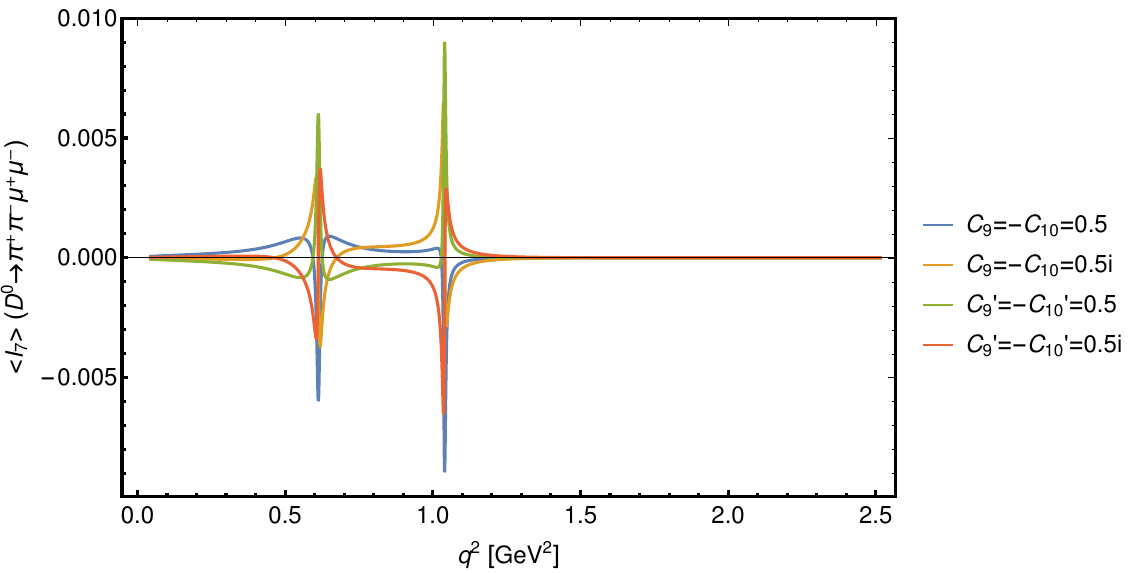}
\caption{Angular observables $\langle I_{5,7}\rangle$ as a function of $q^2$ in $D^0 \to \pi^+ \pi^- \mu^+\mu^-$ as defined by Eq.~\eqref{eq:Ii_ave}. Different NP benchmarks are shown, $\mathcal{C}_9^{(\prime)}=-\,\mathcal{C}_{10}^{(\prime)}=0.5$ with blue (green) and $\mathcal{C}_9^{(\prime)}=-\,\mathcal{C}_{10}^{(\prime)}=0.5\,\text{i}$ with orange (red). The relative strong phase is fixed to $\delta_\rho- \delta_\phi=\pi$. The plots are taken from Ref.~\protect\refcite{deBoer:2018buv}.}
\label{fig:BSMq2}
\end{center}
\end{figure}

Some coefficients $I_i$ can be obtained by symmetric or asymmetric integrations in
the decay angles, as for example the coefficient $I_6$, which is proportional to $\cos \theta_\ell$ and \Afb. This allows for a simplified experimental determination with respect to a full angular analysis. One can define the $p^2$ and $\cos \theta_P$ integrated\break observables as
\begin{align}\label{eq:Ii_ave}
 \langle I_6 \rangle (q^2) & = \frac{1}{\Gamma} \int_{4\, m_P^2}^{\left(m_{D^0}-\sqrt{q^2}\right)^2} d p^2 \int_{-1}^{+1} d \cos \theta_P I_6(q^2,p^2, \cos \theta_P) \, , \\
 \langle I_{5,7} \rangle (q^2) & = \frac{1}{\Gamma} \int_{4\, m_P^2}^{\left(m_{D^0}-\sqrt{q^2}\right)^2} d p^2  \left[ \int_{0}^{+1} d \cos \theta_P  -  \int_{-1}^{0} d \cos \theta_P  \right] I_{5,7}(q^2,p^2, \cos \theta_P) \, ,\nonumber
\end{align}
and focus on the effects from axial-vector NP contributions\cite{deBoer:2018buv}. Figure~\ref{fig:BSMq2}, adopted from Ref.~\refcite{deBoer:2018buv}, shows the integrated $I_{5,7}$ observables as a function of $q^2$ for different NP benchmarks. $I_5$ and $I_6$ have similar BSM-sensitivity. Notice that NP effects are enhanced around the $\rho/\omega$ and the $\phi$ resonances, referred to as \textit{resonance catalyzed}\cite{Fajfer:2012nr}. This has the advantage to allow profiting from the full sample statistics, while NP searches in branching fraction measurements are often limited to restricted $q^2$ ranges which suffer from low statistics.

\subsection{\CP asymmetries}\label{sec:cpasymm} 

\begin{figure}[t]
\centering
\includegraphics[width=0.45\textwidth]{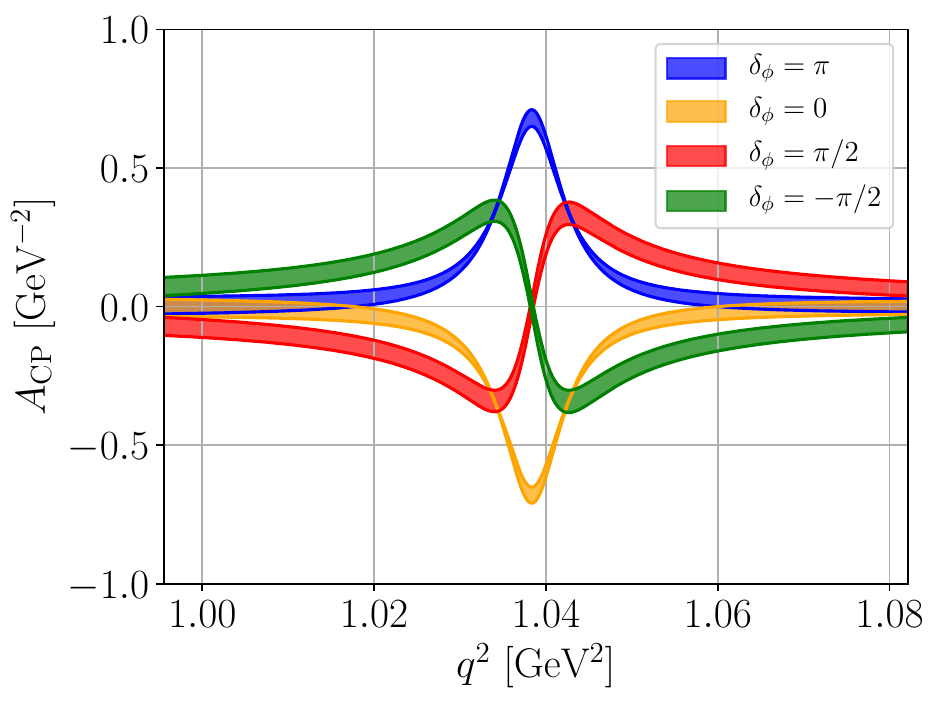}
\includegraphics[width=0.45\textwidth]{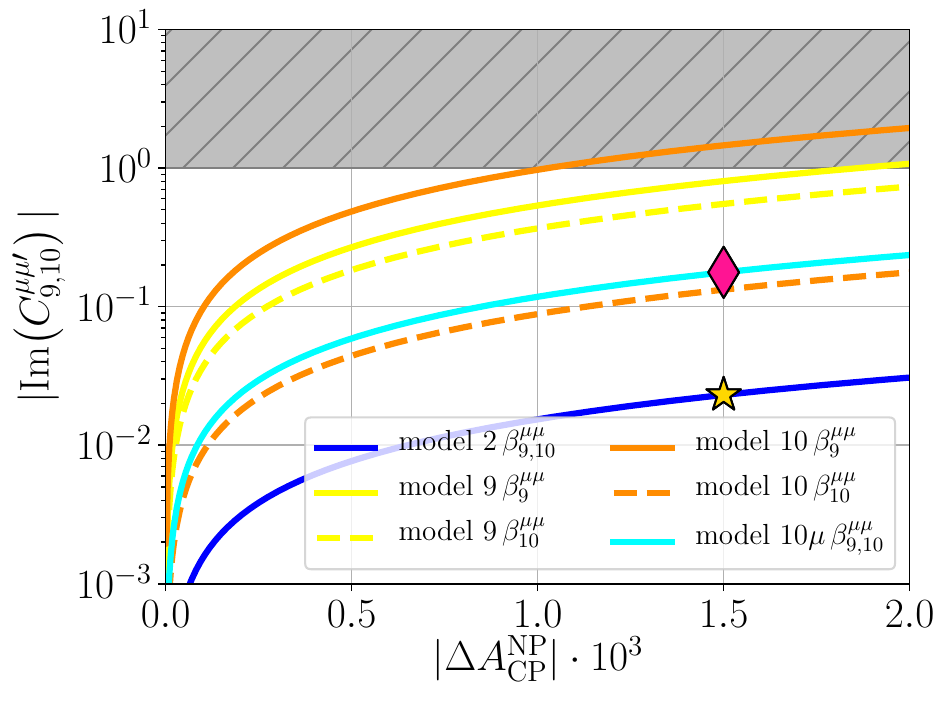}
\caption{Left: $\Acp(D_s^+\to K^+ \mu^+\mu^-)$ with $q^2$ range around the $\phi$ resonance ($q^2\in\left[m_\phi-5\Gamma_\phi,\, m_\phi+5\Gamma_\phi\right]$), exemplary BSM contribution $\mathcal{C}_9^{(\mu)}=0.1\,\exp(\text{i}\,\pi/4)$ and different choices for the unknown strong phase $\delta_\phi$.~\cite{Bause:2019vpr} Right: Magnitude of the imaginary part of $\mathcal{C}_{9,10}^{(\mu)}$ as function of NP contributions to $\Delta \Acp^{\text{NP}}$ in a generic $Z^\prime$-extension~\cite{Bause:2020obd}. NP benchmarks in \Acp (left plot) correspond to the benchmark point (pink diamond) shown in the right plot. The plots are taken from Refs.~\protect\refcite{Bause:2019vpr,Bause:2020obd}. }\label{fig:acpds}
\end{figure}

Since CP-violating effects from SM contributions are CKM-suppressed in rare charm decays, as seen in Sec.~\ref{sec:theoFrame}, studies of CP asymmetries are powerful possibilities to obtain information on NP Wilson coefficients. In general, the CP asymmetry in the dilepton mass distribution is defined as
\begin{equation}\label{eq:ACP}
\Acp (q^2) = {1 \over \Gamma + \overline\Gamma} \left( {\text{d}\Gamma \over \text{d}q^2} - {\text{d}\overline\Gamma \over \text{d}q^2} \right) \,,
\end{equation}
where $\overline\Gamma$ corresponds to the decay rate of the CP-conjugated process. As in Eq.~\eqref{eq:G}, $\Gamma$ and $\bar\Gamma$ denote the $q^2$ integrated decay rates. Similar to the angular observables in $D^0\to P_1 P_2\ell^+\ell^-$ shown in Fig.~\ref{fig:BSMq2}, CP asymmetries generated by NP contributions show maximal signatures in the vicinity of the resonances. The strong phases associated with the resonance contributions are needed to induce nonvanishing asymmetries in the interference with NP amplitudes.
Figure~\ref{fig:acpds} (left) shows the CP asymmetry for the decay mode $D_s^+\to K^+ \mu
^+\mu^-$ in the $q^2$ region around the $\phi$ resonance with exemplary values for the strong phase and the NP contributions. The CP asymmetry can reach the percent level. CP asymmetries of this kind can be defined for any of the decay modes listed in Table~\ref{tab:FFs}, and apparent from Fig.~\ref{fig:acpds}, $q^2$ binning might be necessary to measure a nonzero value.
NP models can only contribute to CP-violating observables in rare charm decays if they extend the SM by additional complex-valued couplings. Correlations between CP asymmetries in rare semileptonic decays and purely hadronic modes, such as the recently observed difference of CP asymmetries in $\Dz \to \Kp\Km$ and $\Dz \to \pip\pim$ decays, $\Delta\Acp$\cite{Aaij:2019kcg}, were studied in Ref.~\refcite{Bause:2020obd} in the context of a generic $Z^\prime$-extension with generation-dependent charges. A SM prediction for $\Delta\Acp$ is not well-established, since different theoretical approaches give predictions differing by a factor $\sim$\,10\cite{Dery:2019ysp,Chala:2019fdb,Buccella:2019kpn,Li:2019hho,Soni:2019xko,Cheng:2019ggx,Khodjamirian:2017zdu,Kagan:2020vri,Nierste:2020eqb,Pich:2019pzg}. In Ref.~\refcite{Bause:2020obd} it was shown that NP contributions of order $10^{-3}$ in $\Delta\Acp$ can induce large imaginary parts of $\mathcal{C}_{9,10}^{(\mu)}$, and vice-versa. This leads to measurable CP asymmetries in semi-leptonic decays at the percent level, as shown in the left of Fig.~\ref{fig:acpds}.

Additional studies on CP asymmetries, mainly within specific BSM models, can be seen in e.g.~Refs.~\refcite{Paul:2011ar,Bigi:2011em,Wang:2014uiz,Delaunay:2012cz,Paul:2012ab} and \refcite{Fajfer:2012nr}.

\subsection{Experimental investigations of angular and CP asymmetries} \label{sec:asymmetriesExp}

\begin{figure}[t]
    \centering
        \includegraphics[width=0.31\textwidth]{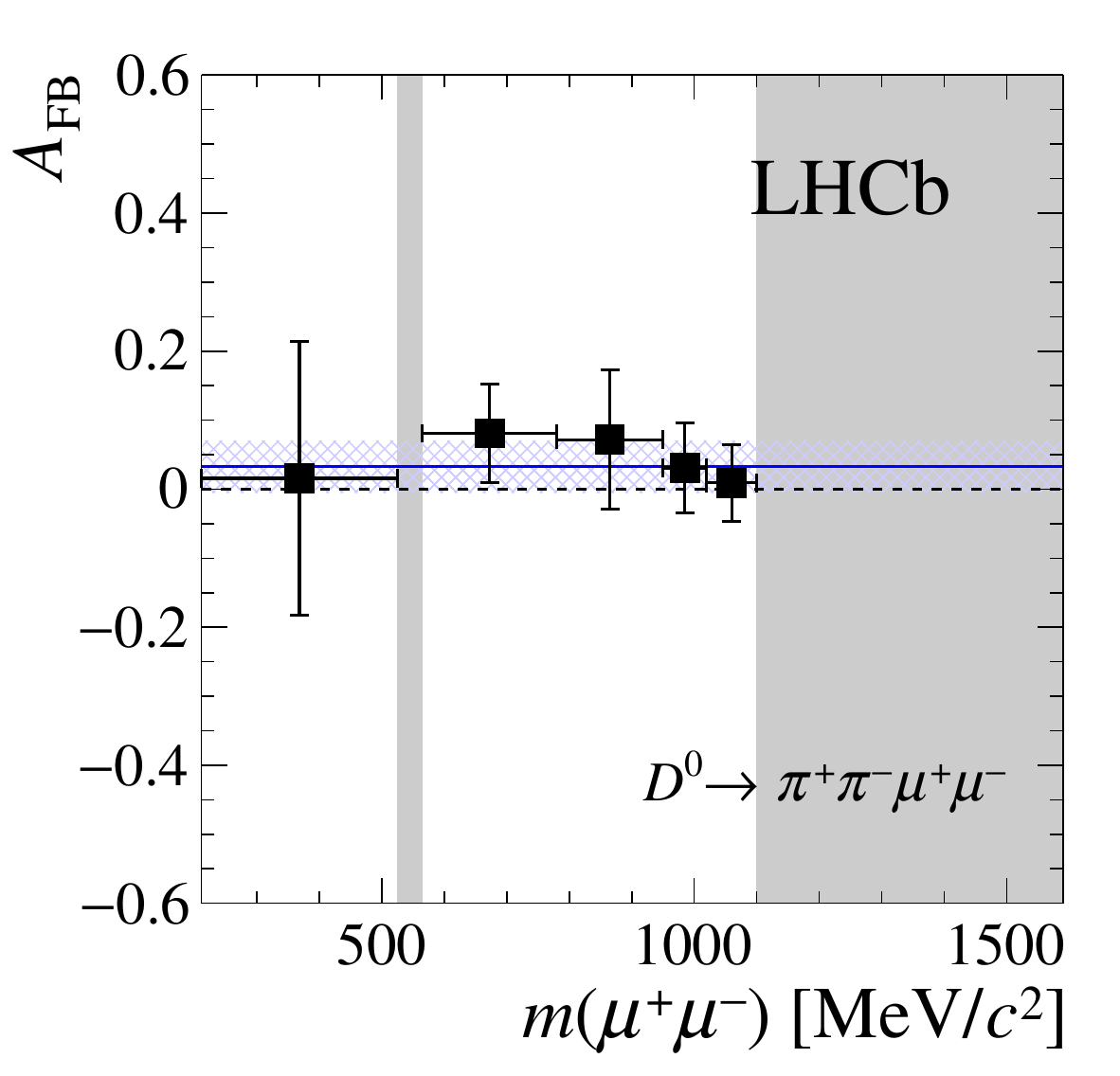}
        \includegraphics[width=0.31\textwidth]{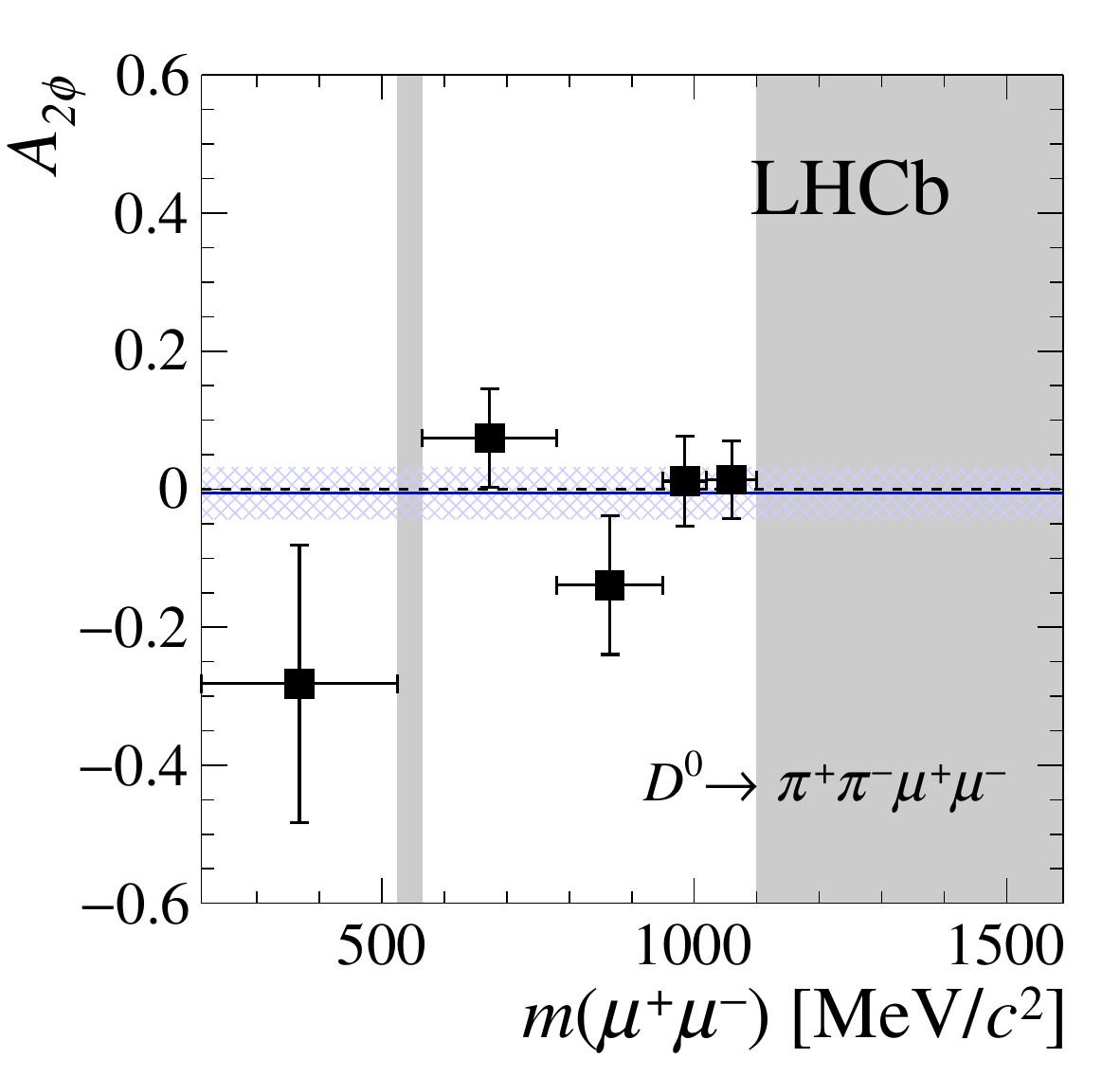}
        \includegraphics[width=0.31\textwidth]{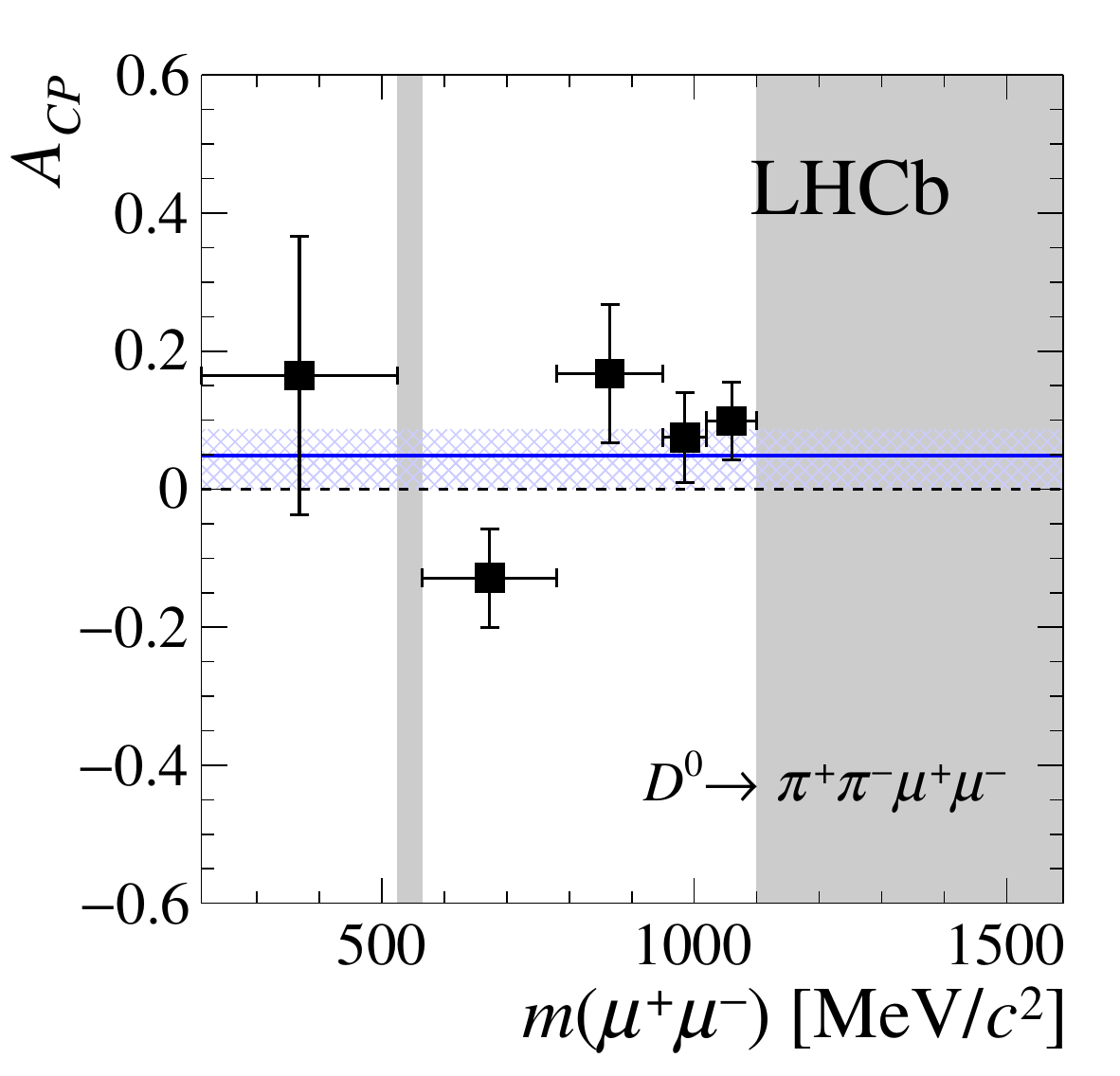} \\
        \includegraphics[width=0.31\textwidth]{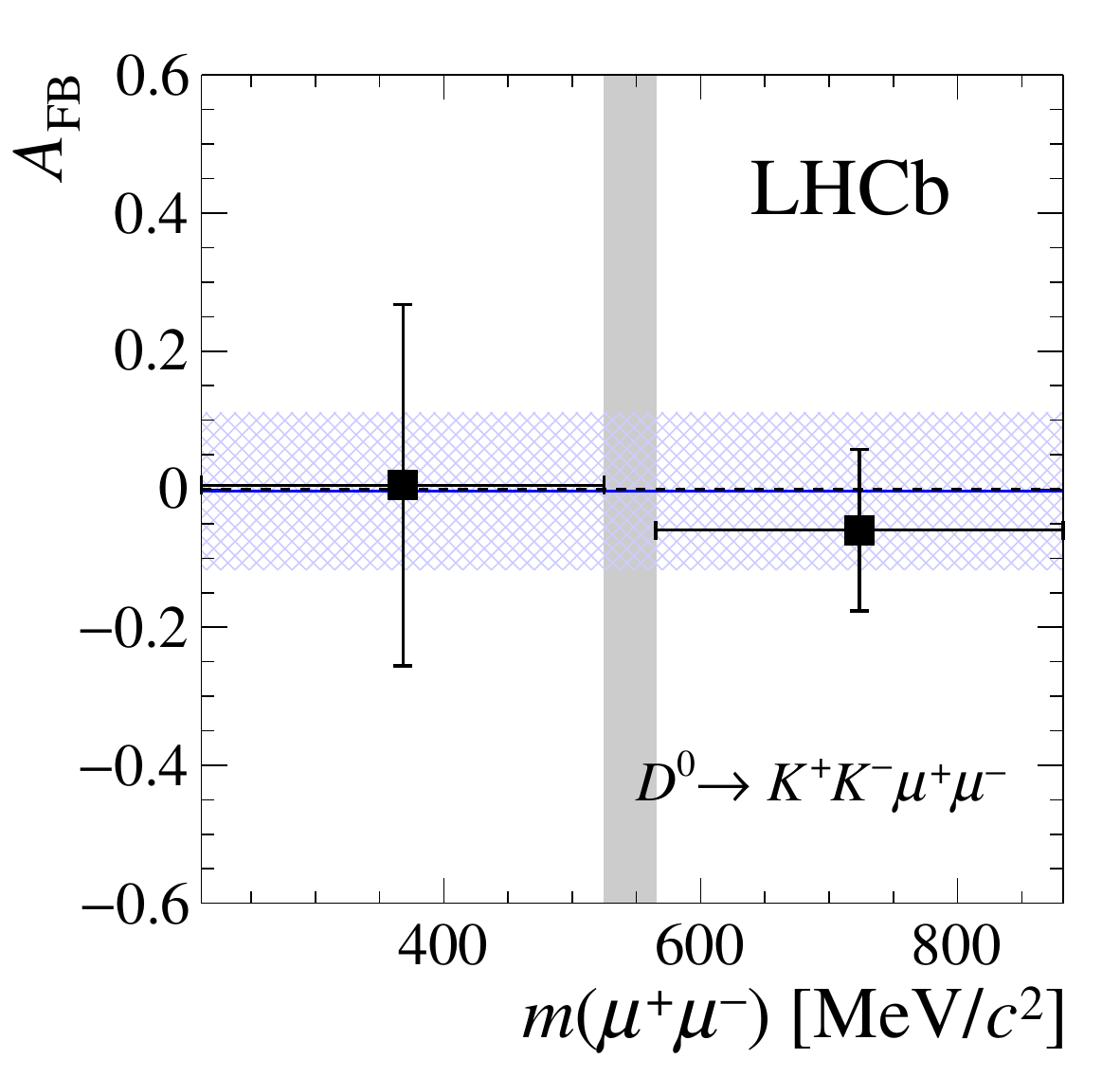}
        \includegraphics[width=0.31\textwidth]{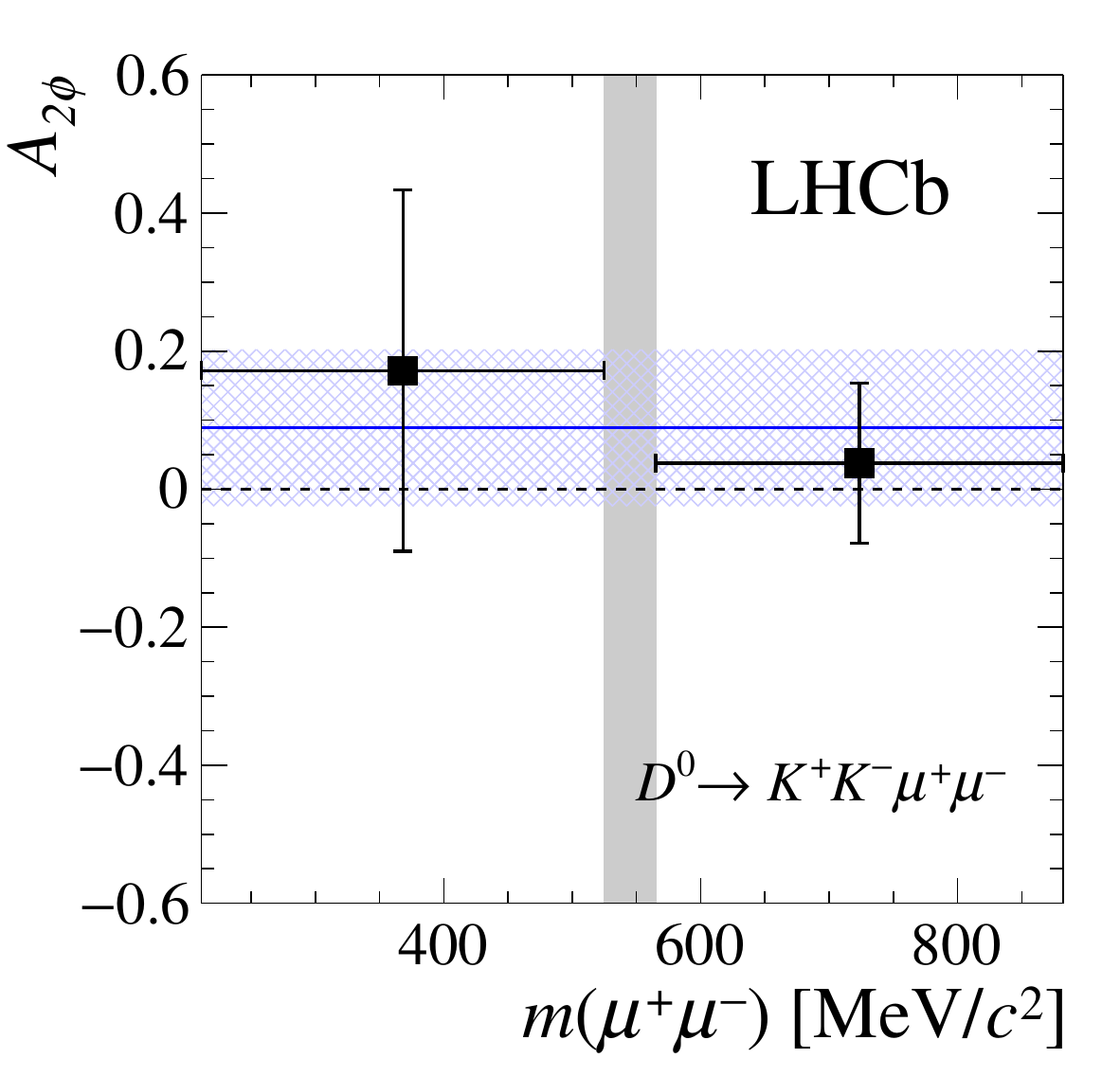}
        \includegraphics[width=0.31\textwidth]{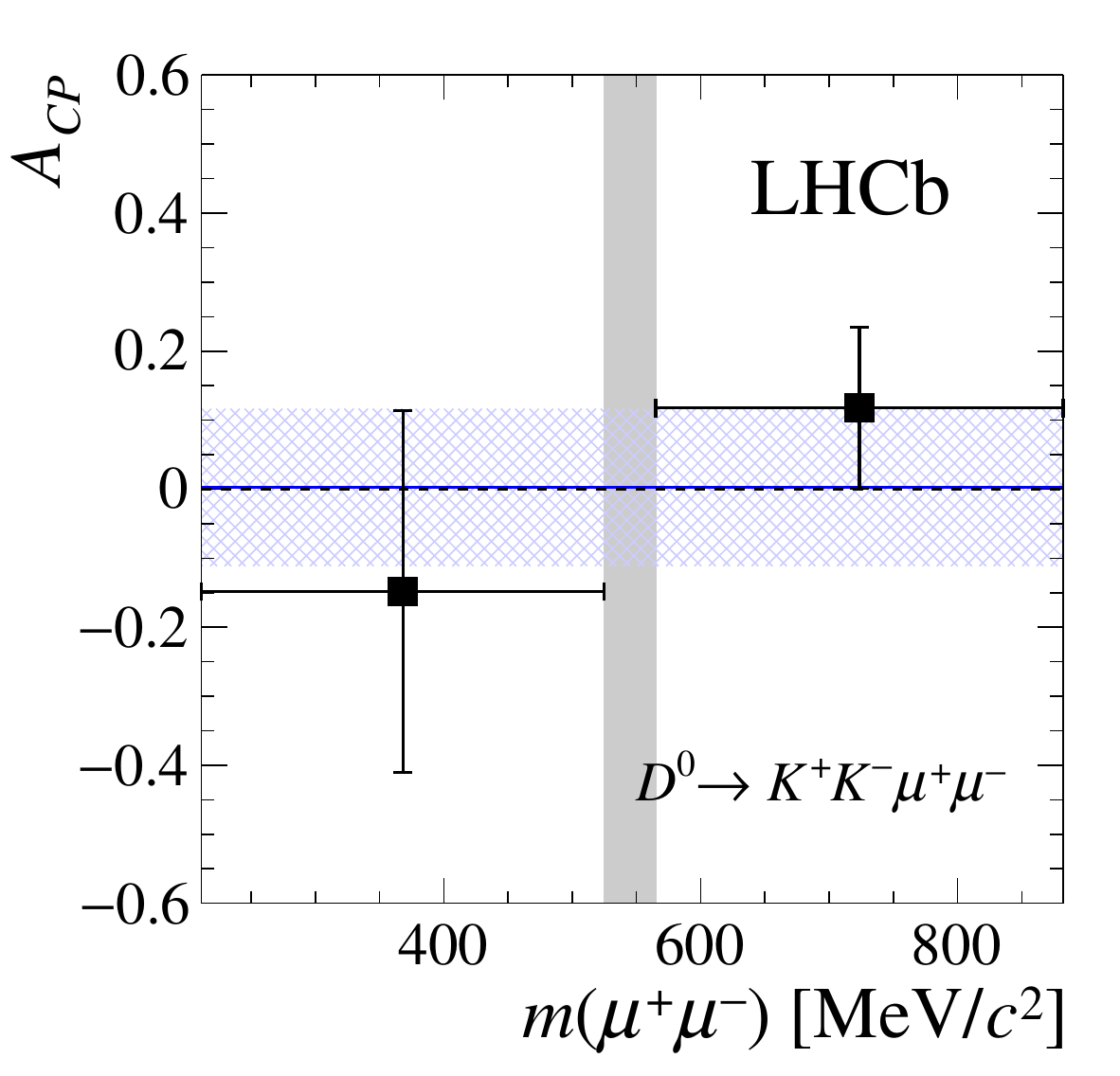} \\
    \caption{Measured forward-backward asymmetry of the dimuon system \Afb, triple-product asymmetry \Aphi and \CP asymmetry \Acp for \Dppmm (top) and \Dkkmm (bottom) decays. Due to a lack of statistics, no measurement has been performed in the grey shaded areas of dimuon mass. The blue line and band show the average and its uncertainty, respectively. Figures are taken from Ref.~\protect\refcite{Aaij:2018fpa}.}
    \label{fig:asymmetries}
\end{figure}
The first measurement of asymmetries in semi-leptonic rare charm decays has been carried out by \lhcb measuring angular and CP asymmetries in \Dppmm and \Dkkmm decays using a dataset corresponding to 5\invfb recorded during the years 2011 to 2016\cite{Aaij:2018fpa}. Among the possible angular asymmetries that can be constructed, the \Afb of the dimuon system, defined as
\begin{equation}
\Afb = \frac{\Gamma(\cos\theta_\mu>0)-\Gamma(\cos\theta_\mu<0)}{\Gamma(\cos\theta_\mu>0)+\Gamma(\cos\theta_\mu<0)}\,, 
\end{equation}
and the triple-product asymmetry, \Aphi, defined as
\begin{equation}
\Aphi = \frac{\Gamma(\sin2\phi>0)-\Gamma(\sin2\phi<0)}{\Gamma(\sin2\phi>0)+\Gamma(\sin2\phi<0)}\,,
\end{equation}
have been investigated\cite{Aaij:2018fpa}. See Sec.~\ref{sec:angularobs} for the definitions of the angles.\footnote{While $\cos\theta_\ell$ is always defined with respect to the positively charged lepton in Sec.~\ref{sec:angularobs}, \lhcb has defined $\cos\theta_\mu$ flavor-dependent as the angle of the positively (negatively) charged muon direction and the direction opposite to the \Dz (\Dzb) meson in the dimuon rest frame.} Comparing Eqs.~\eqref{eq:full} and~\eqref{eq:ang_basis}, \Afb and \Aphi are related to angular observables $I_6$ and $A_9$, respectively, making them SM null tests\cite{deBoer:2018buv,Cappiello:2012vg}. In addition, the CP asymmetry as defined in Eq.~\eqref{eq:ACP} has been measured. Experimentally, the flavor of the \Dz mesons at the moment of their production is determined by selecting neutral charm mesons arising from the decay chain $\Dstarp \to \Dz \pip$, where the charge of the accompanying low-momentum pion unambiguously indicates the flavor of the \Dz mesons. The measured asymmetry is corrected for nuisance charge asymmetries introduced by asymmetric detection efficiencies for positively and negatively charged pions, and for asymmetric production rates of \Dstarp and \Dstarm mesons in \proton\proton collisions. Furthermore, the asymmetries are corrected for phase--space dependent variations of the total reconstruction and selection efficiencies. The investigated asymmetries are found to be\cite{Aaij:2018fpa}
\begin{align}
\begin{split}
\Afb(\Dppmm) &= (\phantom{-}\Afbppmm\pm\AfbppmmStat\pm\AfbppmmSyst)\%~,\\
\Aphi(\Dppmm)&= (\Aphippmm\pm\AphippmmStat\pm\AphippmmSyst)\%~,\\
\Acp(\Dppmm) &= (\phantom{-}\Acpppmm\pm\AcpppmmStat\pm\AcpppmmSyst)\%~,
\end{split}
\end{align}
for the decay mode \Dppmm, and
\begin{align}
\begin{split}
\Afb(\Dkkmm) &= (\Afbkkmm\pm\AfbkkmmStat\pm\AfbkkmmSyst)\%~,\\
\Aphi(\Dkkmm)&= (\Aphikkmm\pm\AphikkmmStat\pm\AphikkmmSyst)\%~,\\
\Acp(\Dkkmm) &= (\Acpkkmm\pm\AcpkkmmStat\pm\AcpkkmmSyst)\%~,
\end{split}
\end{align}
for \Dkkmm. The first uncertainty is statistic and the second one systematic. The asymmetries are consistent with zero and therefore compatible with the SM expectations. The asymmetries have also been investigated as a function of dimuon mass to enhance the sensitivity to NP contributions. The measured asymmetries in bins of dimuon mass are shown in Fig.~\ref{fig:asymmetries}. No dependence of the asymmetries on dimuon mass is found. The precision is limited due to low statistics and reaches a level, where NP predictions start. Conceptually, it is the first measurement of asymmetries in semi-leptonic rare charm decays and we see a large potential in future measurements of a similar kind.

\subsection{Rare radiative charm decays}\label{sec:radiativemodes}

Complementary information with respect to semi-leptonic decays on NP couplings can be obtained from radiative decays. Decays $D^0\to V\gamma$, where $V$ is a light vector meson, are induced by the operators $\mathcal{O}_{1,2}^q$, $\mathcal{O}_7^{(\prime)}$ and $\mathcal{O}_8^{(\prime)}$, see Eq.~\eqref{eq:operators}. In Ref.~\refcite{deBoer:2017que}, branching fractions were estimated using two different methods: First, a QCD-based approach adopted from $B$ physics\cite{Bosch:2001gv,Bosch:2004nd}, where leading power corrections $\sim 1/m_c$ are computed. The approach is limited by large uncertainties on hadronic parameters, which could be constrained by a measurement of the branching fraction of $\Dp\to \rho^+\gamma$ in the future. Second, a hybrid approach combining heavy quark effective theory and chiral Lagrangian\cite{Fajfer:1997bh, Fajfer:1998dv}, which leads to predictions comparable to calculations of previous work, see Refs.~\refcite{Khodjamirian:1995uc} and \refcite{Burdman:1995te} for details. Table~\ref{tab:Dphigamma_DKstar0gamma_branching_ratios} shows the SM predictions for the branching fractions of \Dzpg, \Dzkg and \Dzrg decays as obtained in the two main approaches of Ref.~\refcite{deBoer:2017que}. The SM branching fractions are $\sim$\,$10^{-5}\hbox{--}10^{-4}$ and therefore approximately one to three orders of magnitude above those of resonant-dominated semi-leptonic decays discussed in Secs.~\ref{sec:LU}--\ref{sec:asymmetriesExp}.

CP asymmetries in radiative decays constitute SM null tests. NP models can induce large CP asymmetries up to $\leq 10\%$\cite{deBoer:2017que,Isidori:2012yx,Lyon:2012fk}.
Furthermore, the angular distribution of the photon encodes information on its chirality\cite{Adolph:2018hde,deBoer:2018zhz,Adolph:2020ema,Biswas:2017eyn} and provides additional opportunities to search for NP. Rare radiative charm decays have also been studied in e.g.~Refs.~\refcite{Fajfer:2002gp,Prelovsek:2000xy,Dimou:2012un} and \refcite{Dias:2017nwd}.

\begin{table}[t]
 \tbl{SM branching fractions for $D^0\to V\gamma$ using the approaches discussed in the main text. The QCD based approach scales with poorly constraint hadronic parameters. The table is adapted from Ref.~\protect\refcite{deBoer:2017que}. The experimental results by \belle~\protect\cite{Abdesselam:2016yvr} and \babar~\protect\cite{Aubert:2008ai} are also show in the table, where we added statistical and systematic uncertainties in quadrature. Find details on the measurements in Sect.~\ref{sec:expradiative}. We updated the branching fraction of the normalisation mode used in Ref.~\protect\refcite{Aubert:2008ai} using the most recent value from Ref.~\protect\refcite{10.1093/ptep/ptaa104}.} {
 \resizebox{\textwidth}{!}{ 
 \begin{tabular}{cccc}   \toprule
            &  $\mathcal{B}(\Dzpg)$  $[10^{-5}]$ &  $\mathcal{B}(\Dzkg)$ $[10^{-5}]$ & $\mathcal{B}(\Dzrg)$ $[10^{-5}]$ \\
  \hline
  QCD based approach                    & $0.0074-1.2$   &  $0.11-16$   &     $0.011 - 0.38$   \\
  Hybrid approach                        &  $0.24-2.8$     &  $2.26-46$ &  $0.041-1.17$ \\
  \hline
  Experimental results            &                  &                   &               \\
    \belle \protect\cite{Abdesselam:2016yvr}  &  $2.76\pm0.21$ &  $46.6\pm 3.0$ &  $1.77\pm 0.31$ \\
    \babar\protect\cite{Aubert:2008ai,10.1093/ptep/ptaa104}  &  $2.82\pm0.40$ &  $33.3\pm 3.4$ &  - \\

  \botrule
 \end{tabular} } \label{tab:Dphigamma_DKstar0gamma_branching_ratios} }
\end{table}

\subsubsection{Experimental investigations of radiative decays}\label{sec:expradiative}

The most recent experimental study of the previously discussed decay topologies $\Dz \to V \gamma$ has been made by the Belle collaboration in 2017. They published an analysis\cite{Abdesselam:2016yvr} reporting the first observation of the decay \Dzrg and updated branching fraction measurements of the decays \Dzkg and \Dzpg. The measurement is based on a data set of \ep\en collisions corresponding to an integrated luminosity of 943\invfb and reports the following branching fractions:
\begin{align}\label{eq:radiative_BR}
\begin{split}
\BF(\Dzrg)&=(\BFrg\pm\BFrgStat\pm\BFrgSyst)\BFrgUnit~,\\
\BF(\Dzkg)&=(\BFkg\pm\BFkgStat\pm\BFkgSyst)\BFkgUnit~,\\
\BF(\Dzpg)&=(\BFpg\pm\BFpgStat\pm\BFpgSyst)\BFpgUnit~,
\end{split}
\end{align}
where the uncertainties are statistical and systematic, respectively. The measured branching fractions of the decays \Dzkg and \Dzpg are consistent with previous measurements by \babar\cite{Aubert:2008ai} and the world average. The branching fraction of \Dzrg is slightly larger than predicted in most theoretical calculations\cite{Fajfer:1998dv,Khodjamirian:1995uc,deBoer:2017que}, indicating the poor convergence of the $1/m_c$ and $\alpha_s$ expansion. See Table~\ref{tab:Dphigamma_DKstar0gamma_branching_ratios} for a comparison, where we also add the results of Ref.~\refcite{Aubert:2008ai} for completeness.

In addition, CP asymmetries in these decays have been measured by \belle, representing the first investigation of CP asymmetries in radiative charm decays to date. The flavor of the \Dz meson has been determined by selecting \Dz decays of charged \Dstarp mesons and charge-dependent detection and reconstruction efficiency effects are corrected for using data-driven approaches. The CP asymmetries are measured to be\cite{Abdesselam:2016yvr}
\begin{align}\label{eq:radiative_ACP}
\begin{split}
\Acp(\Dzrg)&=(\phantom{-}\Acprg\pm\AcprgStat\pm\AcprgSyst)\%~,\\
\Acp(\Dzkg)&=(\Acpkg\pm\phantom{0}\AcpkgStat\pm\AcpkgSyst)\%~,\\
\Acp(\Dzpg)&=(\Acppg\pm\phantom{0}\AcppgStat\pm\AcppgSyst)\%~,
\end{split}
\end{align}
where the first and second uncertainties are due to statistic and systematic nature, respectively. At the current level of statistical precision, the measured asymmetries are all compatible with SM expectations and leave large room for NP effects in future measurements. Angular distributions have not yet been studied experimentally\-.

\section{Future prospects}\label{sec:outlook}

We expect significant experimental improvements in the field in the near future, mainly driven by the \lhcb and \belle II collaborations. To date, the \lhcb detector has recorded a data set of \proton\proton collisions corresponding to 9\invfb during 2011--2018. At the moment, the \lhcb detector is undergoing a major upgrade with data taking restart originally planned in 2021. It has been planned to record up to 50 (23)\invfb by 2030 (2024). Due to the COVID-19 situation, currently, a restart in 2022 is planned\cite{Coco}. In parallel, an upgrade phase II is in preparation, led by the ambitious goal to collect up to 300\invfb by 2038. The \belle II detector has started data taking and plans to record a data set of $e^+e^-$ collisions corresponding to 50\invab by the end of 2030\cite{Kou:2018nap,Iijima}.

We expect updated measurements of many searches for rare and forbidden decay modes presented in Sec.~\ref{sec:expSearches} by the current flavor experiments BES III, \belle II and \lhcb. As an example, we summarize future sensitivities of upgrade \lhcb for the selected benchmark channels $D^0 \to \mu^+\mu^-$, $D^+ \to \pip \mumu$ and $\Lambda_c^+ \to \proton \mu^+\mu^-$ as taken from Ref.~\refcite{Bediaga:2018lhg} in Table~\ref{tab:futureLimits}, where conservatively possible improvements in the detector performance have been neglected in the calculations. Using the most recent experimental results\cite{Aaij:2020wyk}, we add projections for the decay channels $D^+ \to \pip e^+e^-$ and $D^+ \to \pip e^+ \mu^-$ to the table by scaling the observed limits to $23\invfb$ and $300\invfb$ of integrated luminosity.\footnote{Assuming the upper limit to scale with the square root of the integrated luminosity.} In particular for semi-leptonic decays involving muons, the projected limits will come close the expected resonant contributions across the full decay phase space. Searches for forbidden decays will reach upper limits below the allowed parameter space of NP modes and therefore have the fantastic potential to find NP or significantly reduce its parameters. Preliminary results\cite{Zhao:2016jna} of the BES III collaboration indicate the potential to make significant contributions to investigations of final states with two electrons. The future limits in Table~\ref{tab:futureLimits} translate into the following model-independent limits on the Wilson coefficients:
\begin{align}
\begin{split}
    &\left|{\mathcal{C}^{(\mu)(\prime)}_{9,\,10}}\right|\lesssim 0.4\,(0.3)~,\quad \left|{\mathcal{C}^{(\mu)}_{T,\,T5}}\right|\lesssim 0.8\,(0.5)~,\quad\,\left|{\mathcal{C}^{(\mu)(\prime)}_{S,P}}\right|\lesssim 0.03\,(0.02) ~,
\end{split}
\end{align}
\begin{align}
    \begin{split}
        &\left|{\mathcal{C}^{(e)(\prime)}_{9,\,10}}\right|\lesssim 2\,(1)~,\quad \left|{\mathcal{C}^{(e)}_{T,\,T5}}\right|\lesssim 2\,(1)~,
    \end{split}
\end{align}
for 23\invfb (300\invfb). Improved bounds on $\mathcal{C}^{(e)}_{S,P}$ require a more precise limit on the branching fraction of $\Dz\to e^+e^-$. Projections for future $\Dz\to e^+e^-$ branching fraction measurements are not available in literature, however, \belle II is expected to be able to significantly improve the current limit.

\begin{table}[t]
 \tbl{Estimated upper limits (UL) of selected rare and forbidden decay modes at \lhcb for future data sets, taken from Ref.~\protect\refcite{Bediaga:2018lhg}. Limits for the decay channels $D^+ \to \pip e^+e^-$ and $D^+ \to \pip e^+ \mu-$ have been obtained by scaling the observed limits taken from Ref.~\protect\refcite{Aaij:2020wyk} to $23\invfb$ and $300\invfb$ of integrated luminosity, assuming the upper limit to scale with the square root of the integrated luminosity.}
 { 
 \centering
 \begin{tabular}{ccc}   \toprule
     Decay channel     &      UL \lhcb extrapolation  & UL \lhcb extrapolation \\
                       &         [$23\invfb$]   &    [$300\invfb$] \\
  \hline
  $D^0 \to \mu^+\mu^-$                &     $\sim 5.9 \times 10^{-10}$   &     $\sim 1.8 \times 10^{-10}$   \\   
  $D^+ \to \pip \mumu$                &     $\sim 1.3 \times 10^{-8} $   &     $\sim 3.7 \times 10^{-9}$   \\
  $\Lambda_c^+ \to \proton \mu^+\mu^-$&        --                        &     $\sim 4.4 \times 10^{-9}$   \\  
  $D^+ \to \pip e^+e^-$               &     $\sim 4.2 \times 10^{-7}$    &     $\sim 1.2 \times 10^{-7}$ \\ 
  $D^+ \to \pip e^+ \mu^-$             &     $\sim 5.5 \times 10^{-8}$    &     $\sim 1.5 \times 10^{-8}$   \\
  \botrule
 \end{tabular} \label{tab:futureLimits}} 
\end{table}

The enormous yields of charm hadrons that will be produced in the future at \lhcb and \belle II will open the door to preform accurate SM null tests. We expect precision measurements of CP asymmetries, angular analyses and tests for lepton universality in semi-leptonic decays of charmed hadrons at the percent level as indicated in Ref.~\refcite{Bediaga:2018lhg}, and even below in resonance-dominated regions of dilepton mass for some semi-leptonic final states. Despite the lack of precise predictions, na\"ive extrapolations of the measured asymmetries in \Dppmm decays\cite{Aaij:2018fpa} at a level of $\sim$\,4\% (see Sec.~\ref{sec:asymmetriesExp}) suggest statistical uncertainties below a percent using the future \lhcb (upgrade phase II) data sets.

\begin{table}[tb]
 \tbl{Estimated uncertainty on \CP asymmetries, \Acp, in rare radiative decay modes at \belle II for future data sets, taken from Ref.\protect\refcite{Kou:2018nap}.}
 {
 \begin{tabular}{cccc}   \toprule
Decay channel     &  $\sigma(\Acp)$ Belle II extrapolation    &$\sigma(\Acp)$ Belle II extrapolation  \\
                  &    [5$\invab$]   &[50$\invab$] \\
\hline
  $\Dzrg$ & $\sim 7\%$   &    $\sim 2\%$   \\
  $\Dzkg$ & $\sim 1\%$   &    $\sim 0.3\%$    \\
  $\Dzpg$ & $\sim 3\%$   &    $\sim 1\%$    \\
  \botrule
 \end{tabular} \label{tab:acpbelle2}} 
\end{table}

We summarize the expected uncertainties on CP asymmetries in rare radiative decays at \belle II in Table \ref{tab:acpbelle2}, which correspond to an improvement of the statistical precision of more than a factor of seven with respect to the current best measurements, which will help to set stringent limits on NP models. We also expect the \belle II collaboration to be capable to investigate angular distributions of rare radiative decays or to study additional topologies, such as $\Dz \to P_1 P_2 \gamma$ ($P_{1,2}=K,\pi$). Both options have recently been suggested by theory\cite{deBoer:2018zhz,deBoer:2017que,Adolph:2020ema}, however, experimental results are not yet available. Efforts including channels with photons and electrons are also expected to be intensified by \lhcb in the future\cite{Bediaga:2018lhg}.

For processes involving LFV with electrons and muons, the future limits in Table~\ref{tab:futureLimits} on $D^+ \to \pip e^+ \mu^-$ are expected to improve the current bounds on Wilson coefficients (see Eq.~\eqref{eq:constraint_D-pimue}) by a factor of 0.5 (0.3) at 23\invfb(300\invfb). Future improvements on the limit of $D^0\to e^\pm\mu^\mp$ are also desired to improve bounds on $K_{S,P}^{(\prime)}$. With the exception of the decay mode $\Dz \to \tau^\pm e^\mp$, information on Wilson coefficients involving $\tau$ leptons are kinematically not accessible in charm hadron decays. Analyses of high-$p_T$ data\cite{Fuentes-Martin:2020lea}, recorded by the ATLAS\cite{Collaboration_2008} and CMS\cite{Chatrchyan:1129810} detectors, will provide additional and complementary constraints. Reference~\refcite{Fuentes-Martin:2020lea} gives estimations for future bounds on $\tau$ Wilson coefficients extrapolating data collected by the ATLAS detector to a data set corresponding to 3\invab (2038)\cite{ATL-PHYS-PUB-2019-005} as
\begin{align}
\begin{split}
    &\left|{\mathcal{C}^{(\tau)(\prime)}_{9,\,10}}\right|\lesssim 3~,\quad \left|{\mathcal{C}^{(\tau)}_{T,\,T5}}\right|\lesssim 2~,\quad\,\left|{\mathcal{C}^{(\tau)(\prime)}_{S,P}}\right|\lesssim 6 ~.
\end{split}
\end{align}

In the longer term, a future circular $e^+e^-$ collider (FCC-ee)\cite{Abada:2019lih} is planned as the next particle collider generation at CERN. Given the benchmark number of $\sim$\,$550 \times 10^9$ produced $c\bar{c}$ pairs at a center-of-mass energy corresponding to the mass of the $Z$ boson\cite{Abada:2019lih}, fantastic ways to probe the SM in rare charm will open. Future tau-charm factories\cite{Zhou:2016qfu,EIDELMAN2015238} operating at energies close to the charm hadron pair production threshold might provide particularly clean collision environments that are well suited for studies of missing energy decay modes, such as dineutrino final states.
First studies of dineutrino decays are already possible today at current $e^+e^-$ colliders, such as BES III and \belle II. Recently, Refs.~\refcite{Bause:2020xzj} and \refcite{Bause:2020auq} proposed the possibility to test charged lepton flavor conservation (cLFC) and LU with dineutrino decays. Using the current bounds on Wilson coefficients from decays of charged leptons, the experimental measurement of a branching fraction of dineutrino decays $h_c\to F\nu\bar\nu$ above $10^{-5}$ would imply a clear sign of cLFC violation and therefore NP. Limits at the order of $10^{-5}$ or less could be possible at the current and future $e^+e^-$ colliders. Further details can be found in Refs.~\refcite{Bause:2020xzj} and \refcite{Bause:2020auq}.

\section{Conclusions} \label{sec:conclusion}

In the past, NP searches in rare decays have mainly focused on $K$ and $B$ systems, and less attention has been devoted to rare charm decays. In this review, we present promising opportunities to test the SM in $|\Delta c| = |\Delta u| = 1$ processes. We discuss the current theoretical status and the most recent experimental measurements. NP searches are currently still possible in branching fraction measurements in restricted regions of the decay phase space, a window that might close soon given the expected sensitivities of current and future flavor experiments. However, we stress and advertise the possibility to define clean null-test observables in resonance-dominated rare and radiative decays. These allow for very clean NP searches with minimal uncertainties from hadronic effects and permit to fully exploit the available statistics of the decays.
Theoretical and experimental exploration of rare charm decays have only started, and the field is expected to play a key role in the future of flavor physics as a complementary testing ground for NP searches. Bright prospects for current and future experimental
flavor facilities such as LHCb, Belle~II, BES~III, the FCC-ee and tau-charm factories
are outlined and open the door to a large and exciting new program in flavor physics. The charm system offers the unique possibility to discover NP in the up-type sector. If the anomalies presently seen in $B$ physics are caused by NP, investigation of rare charm decays will play a crucial and complementary role in the understanding of its origin and nature.

\appendix

\section*{Acknowledgments}
We would like to thank Nico Adolph, Rigo Bause and Gudrun Hiller for useful discussions and enjoyable collaborations. In particular, we are grateful to Gudrun Hiller and Sascha Stahl for reviewing this manuscript. This work is supported by the {\it Studienstiftung des Deutschen Volkes} (MG) and the {\it Bundesministerium f\"ur Bildung und Forschung~--- BMBF} (HG).

\bibliographystyle{ws-mpla}
\bibliography{sample.bib}

\end{document}